\documentclass[useAMS,usenatbib]{mn2e}
\usepackage{mn2ejour}
\usepackage{amssymb}
\usepackage{color}

\usepackage{graphicx}

\title[Cavity in a debris disc]{On the cavity of a debris disc carved by a giant planet}

\author[Zs. Reg\'aly, Z. Dencs, A. Mo\'or, T. Kov\'acs]{Zs. Reg\'aly$^{1}$\thanks{E-mail: regaly@konkoly.hu}, Z. Dencs$^{1,2}$, A. Mo\'or$^{1}$, T. Kov\'acs$^{3,1}$\\
$^1$Konkoly Observatory, Research Centre for Astronomy and Earth Sciences, Hungarian Academy of Sciences,\\\,\,\,\,1121, Budapest, Konkoly Thege Mikl\'os \'ut 15-17, Hungary\\
$^2$Department of Astronomy, E\"otv\"os Lor\'and University, Budapest, Hungary, H-1117 \\
$^3$Institute of Theoretical Physics, E\"otv\"os Lor\'and University, Budapest, Hungary, H-1117
}

\begin{document}


\pagerange{\pageref{firstpage}--\pageref{lastpage}} \pubyear{2017}
 
\maketitle
\label{firstpage}

\begin{abstract}
One possible explanation of the cavity in debris discs is the gravitational perturbation of an embedded giant planet. Planetesimals passing close to a massive body are dynamically stirred resulting in a cleared region known as the chaotic zone. Theory of overlapping mean-motion resonances predicts the width of this cavity. To test whether this cavity is identical to the chaotic zone, we investigate the formation of cavities by means of collisionless \emph{N}-body simulations assuming a $1.25-10$ Jupiter mass planet with eccentricities of $0-0.9$. Synthetic images at millimetre wavelengths are calculated to determine the cavity properties by fitting an ellipse to 14 per cent contour level. Depending on the planetary eccentricity, $e_\mathrm{pl}$, the elliptic cavity wall rotates as the planet orbits with the same ($e_\mathrm{pl}<0.2$) or half ($e_\mathrm{pl}>0.2$) period that of the planet. The cavity centre is offset from the star along the semimajor axis of the planet with a distance of $d=0.1q^{-0.17}e_\mathrm{pl}^{0.5}$ in units of cavity size towards the planet's orbital apocentre, where $q$ is the planet-to-star mass ratio. Pericentre (apocentre) glow develops for $e_\mathrm{pl}<0.05$ ($e_\mathrm{pl}>0.1$), while both are present for $0.05\leq e_\mathrm{pl}\leq0.1$. Empirical formulae are derived for the sizes of the cavities: $\delta a_\mathrm{cav}=2.35q^{0.36}$ and $\delta a_\mathrm{cav}=7.87q^{0.37}e_\mathrm{pl}^{0.38}$ for $e_\mathrm{pl}\leq0.05$ and $e_\mathrm{pl}>0.05$, respectively. The cavity eccentricity, $e_\mathrm{cav}$, equals to that of the planet only for $0.3\leq e_\mathrm{pl}\leq0.6$. A new method based on Atacama Large Millimeter/submillimeter Array observations for estimating the orbital parameters and mass of the planet carving the cavity is also given.
\end{abstract}


\section{Introduction}

Early in the \emph{IRAS} mission three nearby A-type stars, Vega, Fomalhaut, and $\beta$ Pictoris were discovered which show infrared (IR) excess at wavelengths of 25, 60 and 100{\micron} \citep{Aumannetal1984,Gillett1986}. Combining with subsequent spatially resolved scattered light imaging of $\beta$ Pictoris \citep{SmithTerrile1984}, these measurements indicated that the observed excess emission comes from circumstellar discs composed of dust particles larger than the interstellar grains \citep{BackmanParesce1993}. Since, under the influence of stellar radiation forces, such grains are removed on a time-scale much shorter than the age of the star, their presence imply a continuous dust replenishment from collisions/evaporation of larger planetesimals that are unseen for us \citep{BackmanParesce1993}. 

Debris dust is produced when collisions of planetesimals are violent enough to be disruptive, which requires a dynamical stirring of the parent bodies' motion. \citet{KenyonBromley2004} proposed a self-stirring scenario, in which the large (Pluto sized) planetesimals forming via collisional coagulation of smaller bodies, perturb the orbits of neighbouring smaller planetesimals exciting their inclination and eccentricity. \citet{Wyatt2005} and \citet{MustillWyatt2009} proposed an alternative scenario in which a giant planet triggers a collisional cascade by its secular perturbation exciting planetesimals' eccentricities and inclinations.

Passing close to the planet planetesimals feel strong gravitational perturbation. As the eccentricity and inclination of bodies are excited they can scatter out from the system or be accreted by the star. Those planetesimals which enter the instantaneous Hill sphere of the planet can be accreted by the planet itself. Thus, there is a region of unstable orbits on which bodies can not remain in the disc and eventually a cavity  forms in the disc surrounding the planetary orbit called the chaotic zone (see e.g., \citealp{Quillen2006,Suetal2013}). Therefore, the observed dust free inner holes observed in debris discs (see e.g., \citealp{Williamsetal2004}) can be explained as a result of the gravitational perturbation of a giant planet.

Recently, several giant planets were discovered orbiting young (few tens of million years) stars at a large distances ($10-70$\,au), e.g.,  HR\,8799 \citep{Maroisetal2008,Maroisetal2010}, $\beta$\,Pic \citep{Lagrangeetal2009,Lagrangeetal2010}, and HD\,95086 \citep{Rameauetal2013}. Planet(s) and the planetesimal disc can interact with each other in many ways (see e.g., \citealp{Erteletal2012}). Secular perturbations from a inclined or eccentric planet can cause warps or tightly wound spirals in the disc \citep{Roquesetal1994,Mouilletetal1997,BeustMorbidelli2000,Augereauetal2001,Wyatt2005,Dawsonetal2011,Apaietal2015}, and contribute significantly to its stirring and initiation of a collisional cascade far from the planet \citep{Wyatt2008,MustillWyatt2009}. 

Based on the theory of overlapping first-order mean-motion resonances (MMR) \citet{Wisdom1980} derived an analytic formula for the extent of the chaotic zone exterior to the planetary orbit ($a_\mathrm{pl}(1+\delta a_\mathrm{ch})$) for a circular planet and found $\delta a_\mathrm{ch}=1.3q^{2/7}$, where $q$ is the planet-to-star mass ratio.  Considering eccentric planets in numerical simulations \citet{QuillenFaber2006} found that $\delta a_\mathrm{ch}$ is independent of $e_\mathrm{pl}$  for  $q\leq 10^{-3}$. However,  the chaotic zone width was found to be increasing with $e_\mathrm{pl}$  \citep{Bonsoretal2011}, which is due to the widening of MMRs resulting in $\delta a_\mathrm{ch}=1.8(q e_\mathrm{f})^{1/5}$ being valid for $e_\mathrm{pl}\leq e_\mathrm{crit}=2.1q^{1/4}$, where $e_\mathrm{f}$ is the forced eccentricity of particles \citep{MustillWyatt2012,Decketal2013}. For $e_\mathrm{pl}>e_\mathrm{crit}$ numerical simulations showed that $\delta a _\mathrm{ch}\simeq5a_\mathrm{pl,H}$, where $a_\mathrm{pl,H}$ is the planetary Hill radius at its apocentre distance \citep{PearceWyatt2014}. Recently, \citet{MorrisonMalhotra2015} reinvestigate the circular case and found  $\delta a_\mathrm{ch}=1.7q^{0.31}$.

With the advent of Atacama Large Millimeter/submillimeter Array (ALMA) debris discs can be observed with high spatial resolution enabling us to measure the extent and shape of the possible planet carved cavity in these wavelengths. \citet{Erteletal2012} found that  particles trapped in 1:1 MMR of a planet with $e_\mathrm{pl}\leq0.1$ can be detected in millimetre wavelengths. In collisional simulations the co-orbiting 1:1 MMR are also found to be populated for a $1-10\,M_\mathrm{Jup}$ planet on a circular orbit \citep{Chiangetal2009,NesvoldKuchner2015} \citet{Faramazetal2014} modelled the long-term interactions of collisionless particles with  $0.1-1\,M_\mathrm{Jup}$ planet on $e_\mathrm{pl}=0.6$ orbit and found that the 3:2 and 2:1 MMRs inside the chaotic zone are significantly populated.

In this paper, we investigate how the size and shape of the cavity carved by a giant planet is altered by particles in MMRs orbiting inside the chaotic zone. We show that the cavity geometry observable in high-resolution ALMA images can be different from the so-called chaotic zone mainly due to the fact that the cavity shape is always elliptic, whose eccentricity differs from that of the planetary orbit. 

The outline of the paper is the following. The numerical method applied to model the planetesimal-planet interaction and the initial conditions of simulations are given in Section\,2. Section\,3 describes our calculations of synthetic thermal images in the millimetre wavelength using \emph{N}-body simulations. Our findings regarding the properties the cavity for different planetary masses and orbital eccentricities are presented in Section\,4. In Section\,5 a discussion on our results, previous investigations and the observable properties of the planet carved cavity seen on synthetic ALMA images are presented. A new method for the determination of the orbital parameters of a known mass planet based on high-resolution ALMA images is also given in Section\,5. The paper closes with a summary and conclusion in Section\,6.

\section{\emph{N}-body Simulations}

\begin{figure*}
	\begin{center}
	\includegraphics[width=1\columnwidth]{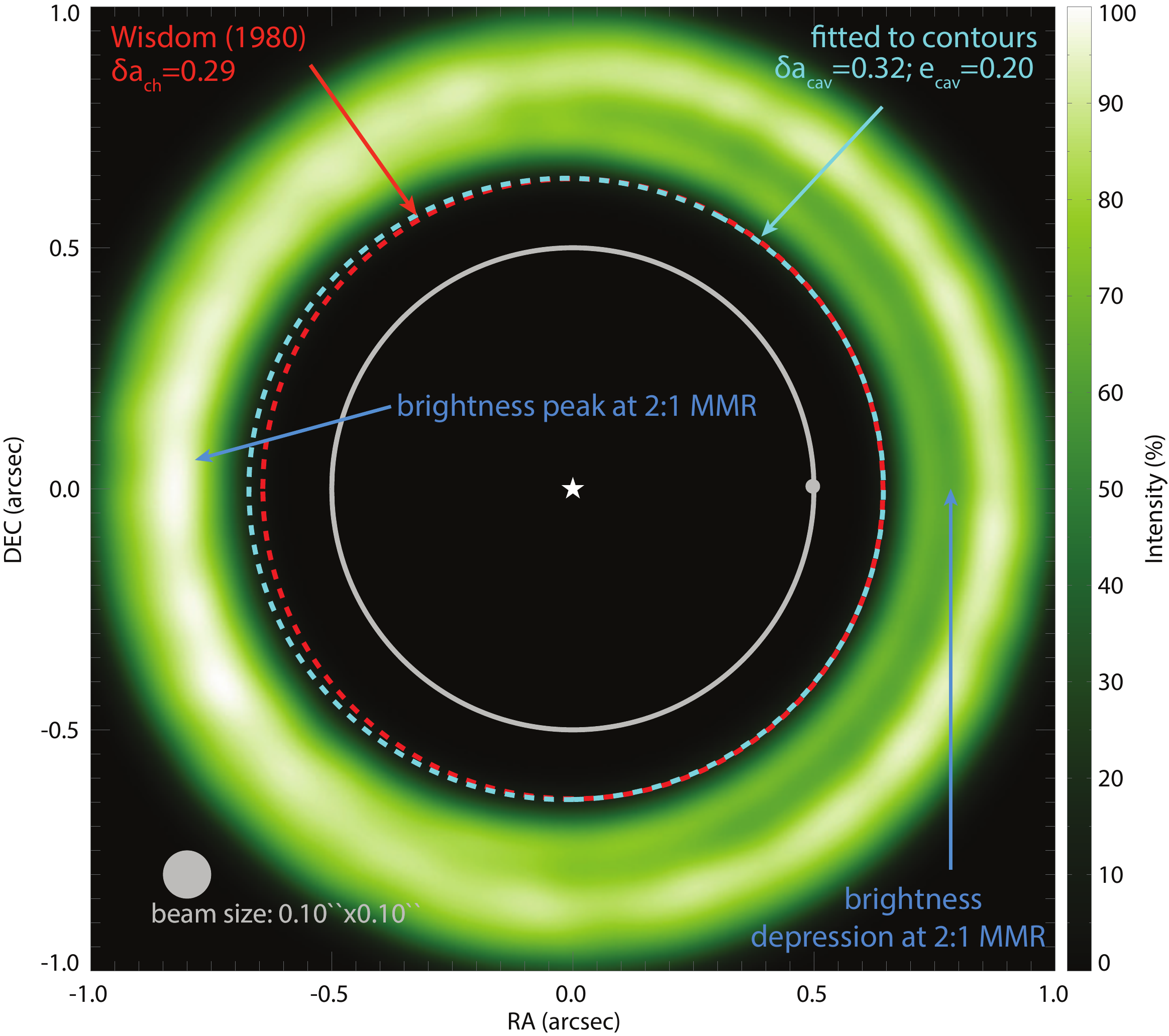}
	\includegraphics[width=1\columnwidth]{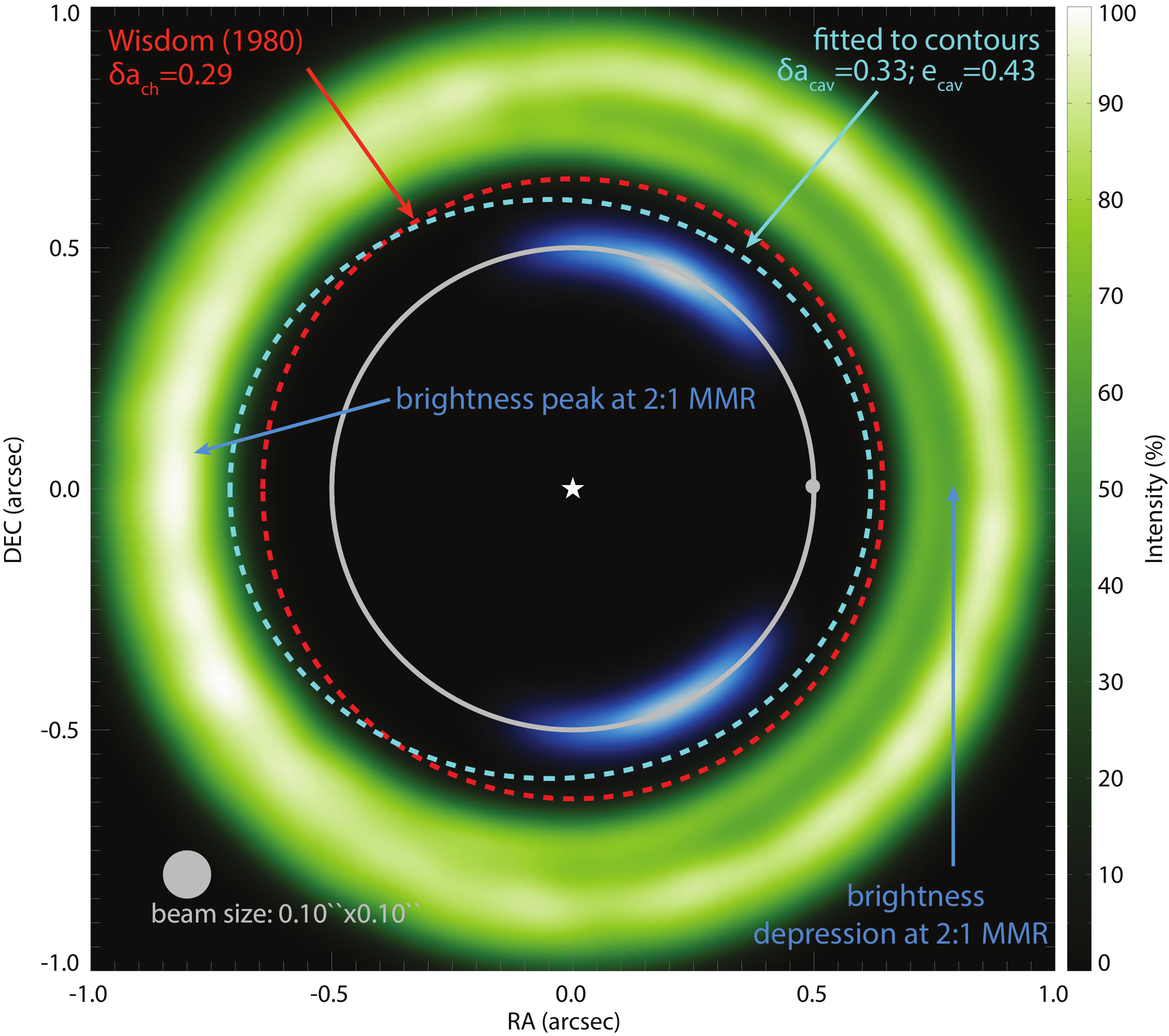}
	\end{center}
	\caption{Calibration of the chaotic zone by fitting an ellipse to the dust emission (normalized to the maximum intensity and convolved with a $0.1{\arcsec}\times0.1{\arcsec}$ Gaussian beam) at 1.4\,mm for $q=5\times 10^{-3}$ and $e_\mathrm{pl}=0$ cold disc model. Dust particles orbiting in 1:1 (blue) MMR are neglected for the calibration process (left) and included for the general fitting procedure (right). Grey circle represents the planetary orbit. To emphasize particles trapped in 1:1 MMR their emission is artificially strengthened. Note that during the cavity fitting process no artificial intensity strengthening was applied.}
	\label{fig:calibration}
\end{figure*}

While optical and IR measurements predominantly trace micrometer-sized grains whose spatial distribution are significantly affected by the stellar radiation pressure, observations at millimetre wavelengths probe rather those millimetre-sized particles that are largely insensitive to radiation forces and thereby can serve as proxy for their parent planetesimals.  To investigate the formation of a cavity by an embedded giant planet in debris discs, we run \emph{N}-body simulations modelling the gravitational perturbation of a giant planet on the planetesimal disc.

We apply the restricted three body approach, i.e., the star and giant planet gravitationally interact with each other and perturb the orbits of planetesimals. {In this approach the orbits of massive bodies are not perturbed by the planetesimals}. We use a Graphical Processing Unit (GPU) based direct \emph{N}-body integrator {\sc HIPERION}\footnote{http://www.konkoly.hu/staff/regaly/research/hiperion.html.}, utilizing sixth-order Hermite scheme with double precision \citep{MakinoAarseth1992,NitadoriMakino2008}.  The {\sc HIPERION} code calculates all steps of the gravitational interactions and the solutions of the equation of motion on a GPU, i.e. no host-to-device communication is required. The simulations were performed on NVIDIA Tesla 2075 and K40 GPUs with a mean flop rate 135\,GFLOPS\,$s^{-1}$.

The  sixth-order Hermite scheme is found to be the most effective (taking into account the mean iteration time for a given precision using various order schemes) for our purposes. Initially $N=5\times10^5$ massless particles  are in the computational domain, whose number decrease to about $2.25\times10^5$ by the end of each simulation because of scattering out or accreting by the star or planet. The large number of particles used in the simulations enables us to create synthetic millimetre images with low Poisson noise.

The planet-to-star mass ratio ($q$) is in the range of $1.25\times10^{-3}-10\times10^{-3}$ (corresponding to 1.25-10 Jupiter mass for a Solar mass star in the centre). We investigate models with different planetary eccentricity ($e_\mathrm{pl}$) in the range of $0-0.9$.  To model the clearing of the cavity formed around the planetary orbit for wide range of $e_\mathrm{pl}$ the semimajor axis of planet is set such that the planetary apocentre is fixed at 1, i.e., $a_\mathrm{pl}=1/(1+e_\mathrm{pl})$ in our units.

The semimajor axis of particles initially follow uniform distribution in the range of $0.8\,a_\mathrm{pl}-2\,a_\mathrm{pl}(1+e_\mathrm{pl})$ and are on Keplerian orbits around the barycentre. The ascending node and the argument of periapses of particles are uniformly distributed in the range $0-2\mathrm{\pi}$.
 
We investigate collisionally excited planetesimal configurations for which cases the eccentricity  and inclination of  particles follow Rayleigh distribution \citep{Lissauer1993}.  In order to investigate the effect of the planetesimals' initial eccentricity and inclination distribution on the formation of the planetary cavity, we study dynamically cold and hot discs. In these cases the planetesimals are initially distributed in a vertically slim and thick disc, respectively. 

It is known that the initial mean square values of the eccentricity, \textless$e_\mathrm{plms}^2$\textgreater, and inclination, \textless$i_\mathrm{plms}^2$\textgreater, of planetesimals are such that \textless$e_\mathrm{plms}^2$\textgreater$=$\textless$2i_\mathrm{plms}^2$\textgreater  based on \emph{N}-body simulations \citep{IdaMakino1992}. Thus, assuming that the initial disc configuration is the result of collisional excitations, the planetesimals' initial eccentricity and inclination are assumed to be \textless$e_\mathrm{plms}^2$\textgreater$=0.01$ and \textless$i_\mathrm{plms}^{2}$\textgreater$=0.005$ for cold, and \textless$e_\mathrm{plms}^2$\textgreater$=0.05$ \textless$i_\mathrm{plms}^{2}$\textgreater$=0.025$ for hot disc models.

Adaptive shared time-step is used with second-order Aarseth scheme \citep{PressSpergel1988} with $\eta=0.015-0.02$ (higher planetary eccentricity requires smaller $\eta$ to limit the calculation error to a certain value). The total energy of the system is conserved with accuracy of $dE/E_0<10^{-9}$ by the end of the simulations in all models.

Adaptive time-step method requires to remove particles -- which gain large acceleration due to close encounters with the massive objects -- from the computational domain. Therefore, the planet can accrete particles which enter its instantaneous Hill sphere ($R_\mathrm{H}=d_\mathrm{pl}(q/3)^{1/3}$, $d_\mathrm{pl}$ being the  stellar distance).\footnote{Note that the size and shape of the cavity carved by the planet is found to be the same if planetary accretion is not allowed.} Additionally, particles approaching the star closer than $0.1\,a_\mathrm{pl}$ or go beyond $30\,a_\mathrm{pl}$ are also removed from the computational domain. With these conditions particles with high eccentricities (about 0.8) still remain inside the computational domain.

In a certain region along the planetary orbit periodic orbits do not exist, only conditionally periodic or chaotic ones present, due to the gravitational perturbation of the planet. As a result, this region, the so-called chaotic zone, is largely emptied of particles. Assuming that the chaotic zone edge is about $a_\mathrm{ch}\simeq2a_\mathrm{pl}$ distance, the time-scale of secular perturbation at a distance of the chaotic zone edge is $t_\mathrm{sec}=  2 \pi (1-e_\mathrm{pl}^2)^{3/2}/[\sqrt{G M_*/a_\mathrm{ch}^3} (3q/4)(a_\mathrm{ch}/a_\mathrm{pl})^2]$  corresponding to $N_\mathrm{sec}\simeq 113(1-e_\mathrm{pl}^2)^{-3/2}$ number of orbits \citep{Kaula1962,Heppenheimer1978,MustillWyatt2009}. Therefore, we ran simulations by $5\times10^4$ and $2\times10^5$  orbits for $e_\mathrm{pl}\leq0.5$ and $e_\mathrm{pl}\geq0.6$, respectively, corresponding to $\sim500-2000N_\mathrm{sec}$.

\section{Measurements on cavity}
\label{sec:dust-em}

Assuming that the spatial distribution of millimetre-sized dust particles  follows that of their parent planetesimals \citep{Wyatt2006}, the outer edge of the planet carved cavity can be determined based on the thermal emission of dust particles at millimetre wavelengths. First, the \emph{N}-body simulations are spatially scaled up  by 50, thus planetary apocentre distance corresponds to 50\,au. We use DUSTMAP\footnote{'Synthesizes images of simulated debris discs', {\sc DUSTMAP 3.1.2} is developed by Christopher C. Stark. The package  can be downloaded from http://www.starkspace.com/code/} to calculate the emission of dust particles assuming that the disc is at 100\,pc distance. The stellar parameters required for calculating the disc's thermal emission correspond to a 20\,Myr old Solar type star: $M_*=1\,M_\odot$, $T_*=4580$\,K$, L_*=0.48\,L_\odot$, and $\log g=4.5$ \citep{Siessetal2000}.  

We chose the stellar age to be consistent with the age required to clear the cavity by the giant planet even for the largest applied planetary eccentricity (see details in Section\,\ref{sect:conv_state}). The absorption and emission properties  of dust particles  correspond to that of astronomical silicates \citep{DraineLee1984}. The synthetic images are calculated at a wavelength of $\lambda=1.3$\,mm with resolution of  $512\times512$ pixels corresponding to spatial resolution of $0.002{\arcsec}$. To mimic continuum observations at millimetre wavelengths by ALMA a Gaussian convolution is applied with beam size of $0.1{\arcsec}\times0.1{\arcsec}$. As a last step, images are normalized by their maximum pixel intensity.

The size of the outer planetary chaotic zone can be given as $a_\mathrm{ch}=a_\mathrm{pl}(1+\delta a_\mathrm{ch})$, where $\delta a_\mathrm{ch}$ is the width of the chaotic zone. Based on the theory of overlapping resonant orbits $\delta a_\mathrm{ch}\simeq1.3q^{2/7}a_\mathrm{pl}$, which gives $\delta a_\mathrm{ch}=0.29$ for our $q=5\times10^{-3}$ circular model \citep{Wisdom1980}. Since the chaotic zone can be populated by resonant particles, the size of the cavity identified on observed images is different.  Therefore, we determine the cavity by selecting appropriate contour lines of the emission (close to the cavity edge) and fit an ellipse based on the method of \citet{Markwardt2009}. The cavity size and shape can be described by the semimajor axis, $a_\mathrm{cav}=a_\mathrm{pl}(1+\delta a_\mathrm{cav})$, and the eccentricity, $e_\mathrm{cav}$, of the best fitting ellipse. 

The appropriate intensity range for the contour level is determined such that the theoretical prediction of the chaotic zone width and our measurements of the cavity size agree, i.e. $\delta a_{cav}\simeq\delta a_\mathrm{ch}$. Since particles in MMRs are neglected in the overlapping resonance theory, the emission of dust particles in the co-orbital (1:1) MMR for the $e_\mathrm{pl}=0$ model is artificially removed during the calibration process.
 
Left hand panel of Fig.\,\ref{fig:calibration} shows the result of our calibration process for $q=5\times 10^{-3}$ and $e_\mathrm{pl}=0$ cold disc model.  The cavity width  is $\delta a_\mathrm{cav}=0.3$, if we select contours that have intensities between 14\% and 15\%. Interestingly the cavity is not circular but has an eccentricity of $e_\mathrm{cav}=0.21$ even for a circular planetary orbit. As particles in MMRs may contribute to the millimetre-emission with a significant level, they must be taken into account when calculating the cavity size, for which case we found $\delta a_\mathrm{cav}=0.37$ and $e_\mathrm{cav}\simeq0.44$  (right hand panel of Fig.\,\ref{fig:calibration}). 

Testing the calibration process with different emission wavelengths in the range of $870\mu \mathrm{m}-3\rm mm$ we did not find any change in the size and shape of the planet carved cavity.  We also tested the calibration process against the stellar luminosity and mass by setting the stellar age between $6\,\mathrm{Myr}-1\,\mathrm{Gyr}$ and $0.5\,M_\odot-2\,M_\odot$, respectively. We found no significant dependence of the cavity size and shape on the stellar luminosity and mass, which can be explained by that the millimetre-emission is concentrated to a narrow belt and we normalize the synthetic images.

In the followings to determine the planet carved cavity size and shape, contour lines in the intensity range of $14\%-15\%$ per cent (taking into account the emission of particles in MMRs) are selected and fitted by an ellipse.

\section{Results}

\subsection{Converged equilibrium state of cavity}
\label{sect:conv_state}

\begin{figure}
	\includegraphics[width=\columnwidth]{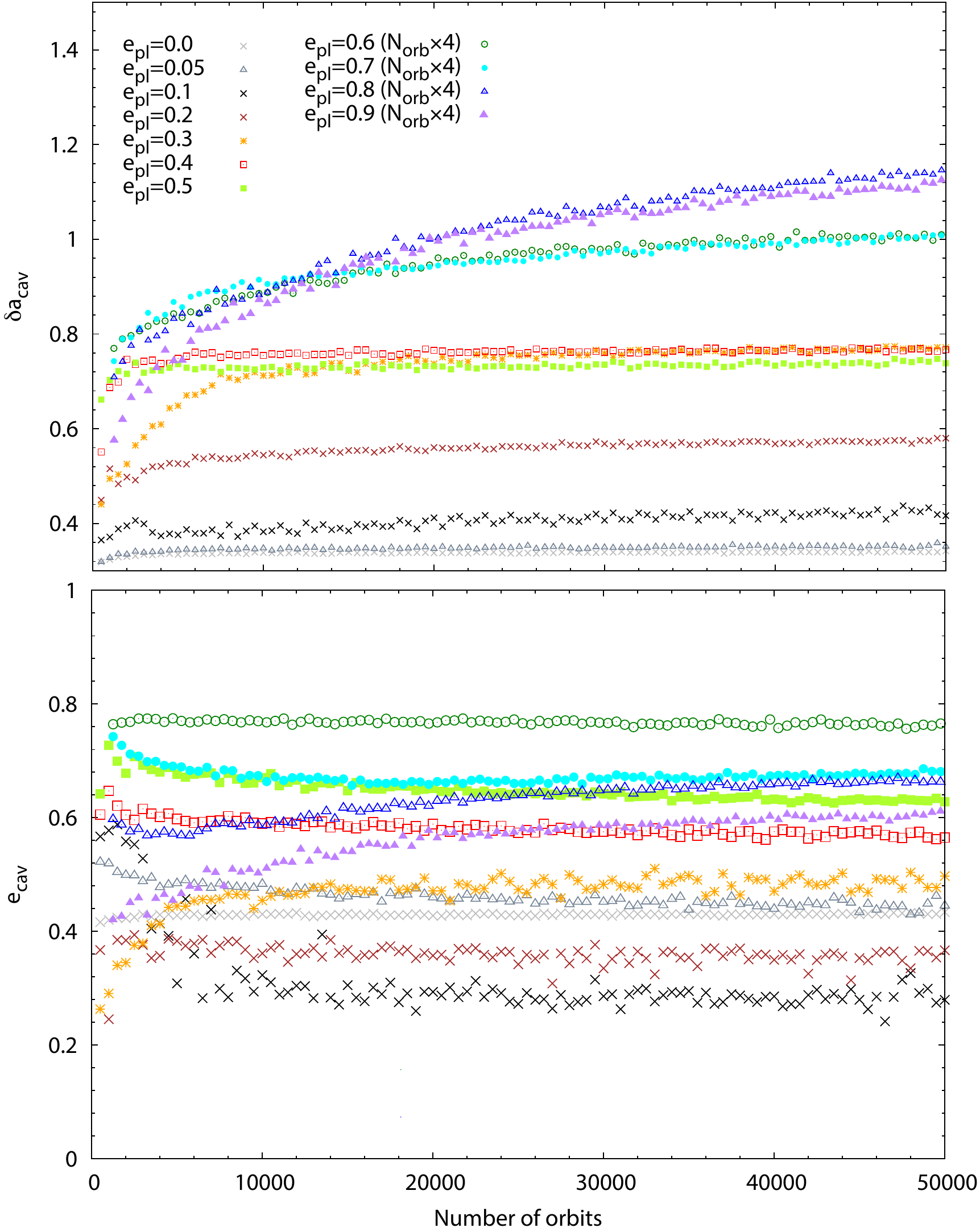}
	\caption{The size (top panel) and eccentricity  (bottom panel) of the cavity as a function of the number of planetary orbits for $q=5\times10^{-3}$ cold disc models. Note that the number of orbits must be multiplied by $4$ for $e_\mathrm{pl}\geq0.6$ models.}
	\label{fig:a_ch-saturation}
\end{figure}

In order to check that the cavity is in a converged equilibrium state at the end of simulations $\delta a_\mathrm{cav}$ and $e_\mathrm{cav}$ are determined at every 500th planetary orbits when the planet is at apocentre. Top panel of Fig.\,\ref{fig:a_ch-saturation} shows $\delta a_\mathrm{cav}$ against the number of planetary orbits for different $e_\mathrm{pl}$ cold disc models. For moderate eccentricities ($e_\mathrm{pl}\leq0.5$) $\delta a_\mathrm{cav}$ is saturated by $5\times10^4$ orbits, however, for large planetary eccentricities ($e_\mathrm{pl}\geq0.6$)  $\delta a_\mathrm{cav}$ saturates only by $2\times10^5$ orbits.

Bottom panel of Fig.\,\ref{fig:a_ch-saturation} shows $e_\mathrm{cav}$ against the number of planetary orbits. $e_\mathrm{cav}$ clearly shows saturation in all models by $10^4$ orbits for $e_\mathrm{pl}\leq0.7$ and by $10^5$ for $e_\mathrm{pl}\geq0.7$. $e_\mathrm{cav}$ is found to be greater than $e_\mathrm{pl}$ for $e_\mathrm{pl}\leq0.6$, while $e_\mathrm{cav}$ always lower than $e_\mathrm{pl}$ for  $e_\mathrm{pl}\geq0.7$. For hot disc models, the same behaviour of growth rates of $\delta a_\mathrm{cav}$ and $e_\mathrm{cav}$ were found.

\subsection{Role of resonant particles}

\begin{table}
\begin{center}
\caption{Most populated MMR inside the chaotic zone given by \citet{PearceWyatt2014} in hot and cold disc models with 5 Jupiter mass planet.}
\label{tbl-1}
\begin{tabular}{clll}
\hline
$e_\mathrm{pl}$ & 
 MMR\,1 &
 MMR\,2 &
 MMR\,3
\\
\hline
0	& 1:1 & 3:2$^b$ & $-$             \\
0.05	& 1:1 & 3:2$^b$ & $-$                                        \\
0.1	& 1:1 & 3:2 & $-$                                   \\
0.2	& 1:1 & 3:2 & 2:1                               \\
\hline
0.3	& 2:1 & 5:2$^a$ &  $-$                                      \\
0.4	& 2:1 & 5:2 &  $-$                                        \\
0.5	& 2:1 & 5:2 & $-$                                       \\
0.6	& 2:1 & 5:2 & 3:1$^a$                                \\
0.7	& 2:1 & 3:1 & $-$                               \\
\hline
0.8	& 3:1 & 4:1$^a$ &  $-$                                       \\
\hline
0.9	& 4:1 & $-$ & $-$             \\
\hline
\end{tabular}

{$^a$}{\,Populated only for cold disc models.}
{$^b$}{\,MMR is not detached from the disc.}
\end{center}
\end{table}

\begin{figure}
	\includegraphics[width=\columnwidth]{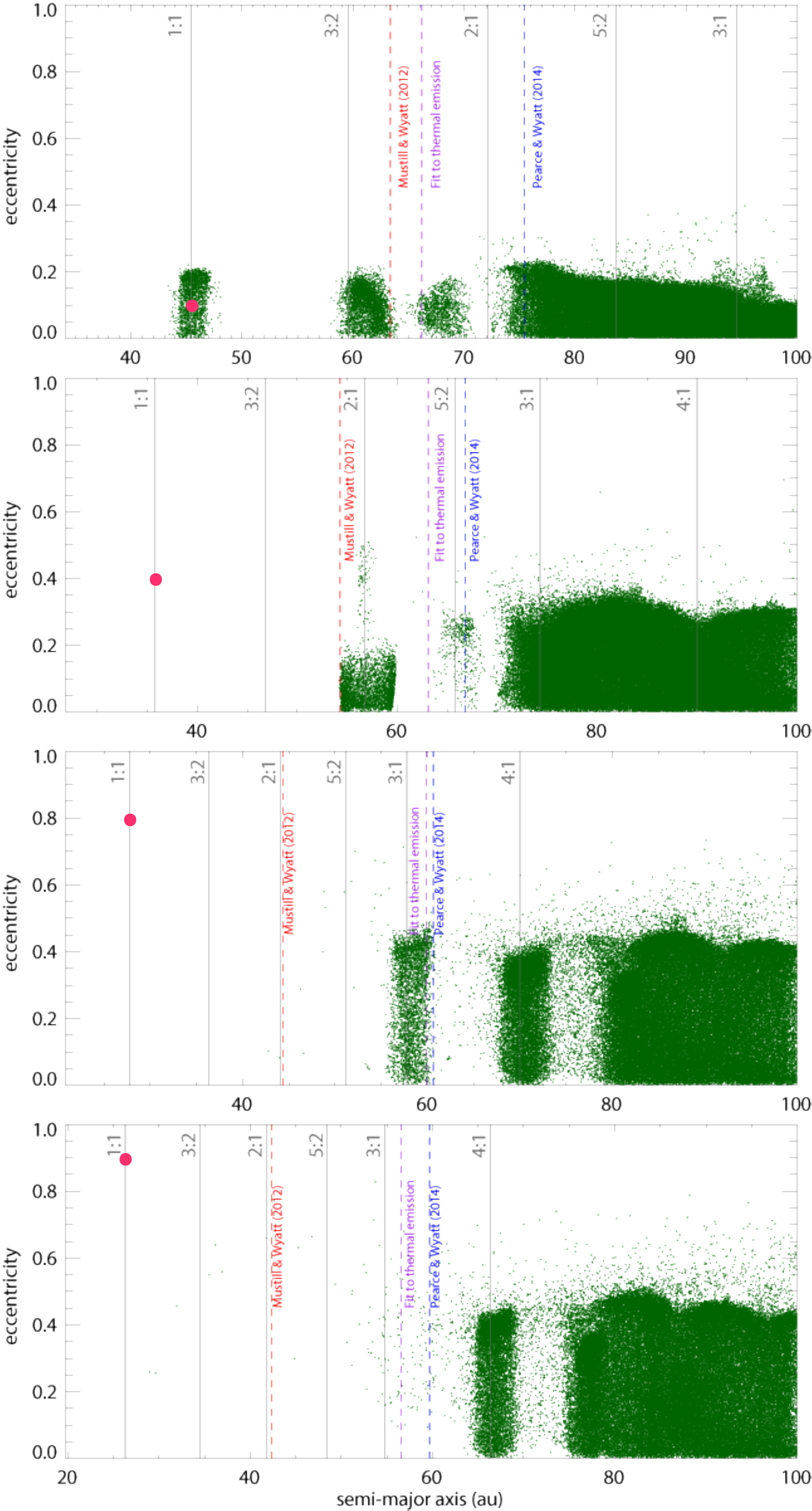}
	\caption{Semimajor axis versus eccentricity of particles for $q=5\times10^{-3}$ and $e_\mathrm{pl}=0.1,0.4, 0.8$, and $0.9$ cold disc models ($a_\mathrm{pl}$ decreases with  increasing $e_\mathrm{pl}$). Positions of MMRs are shown with grey lines. The outer chaotic zone size given by \citet{MustillWyatt2012} and \citet{PearceWyatt2014} are shown with red and blue dashed lines, respectively. The cavity sizes derived by our thermal emission method are shown with purple dashed lines. The giant planet is represented by red filled circle.}
	\label{fig:a-e}
\end{figure}

\begin{figure*}
	\begin{center}
	\includegraphics[width=2\columnwidth]{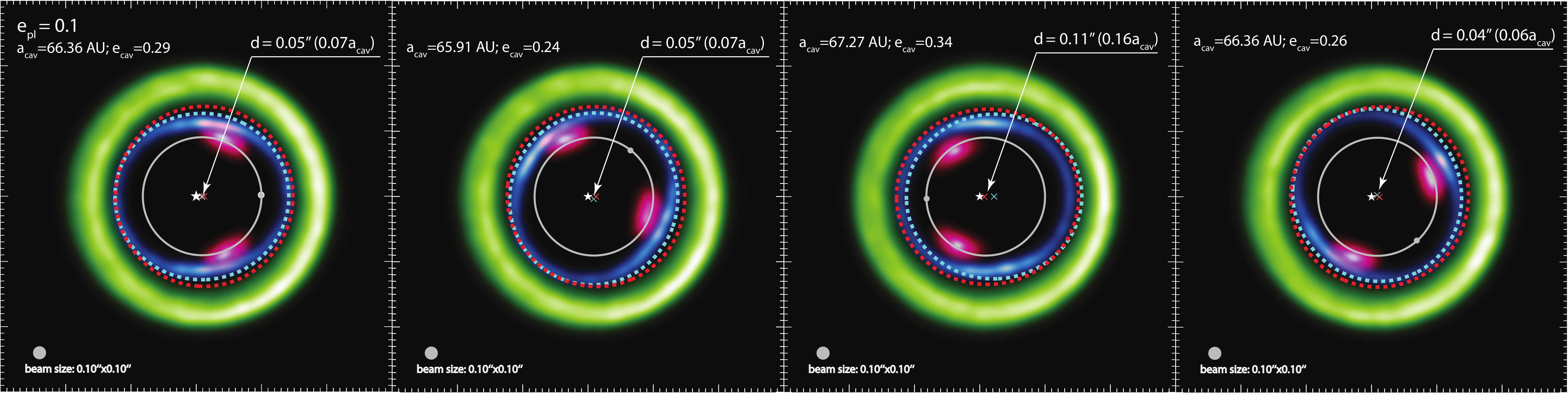}
	\includegraphics[width=2\columnwidth]{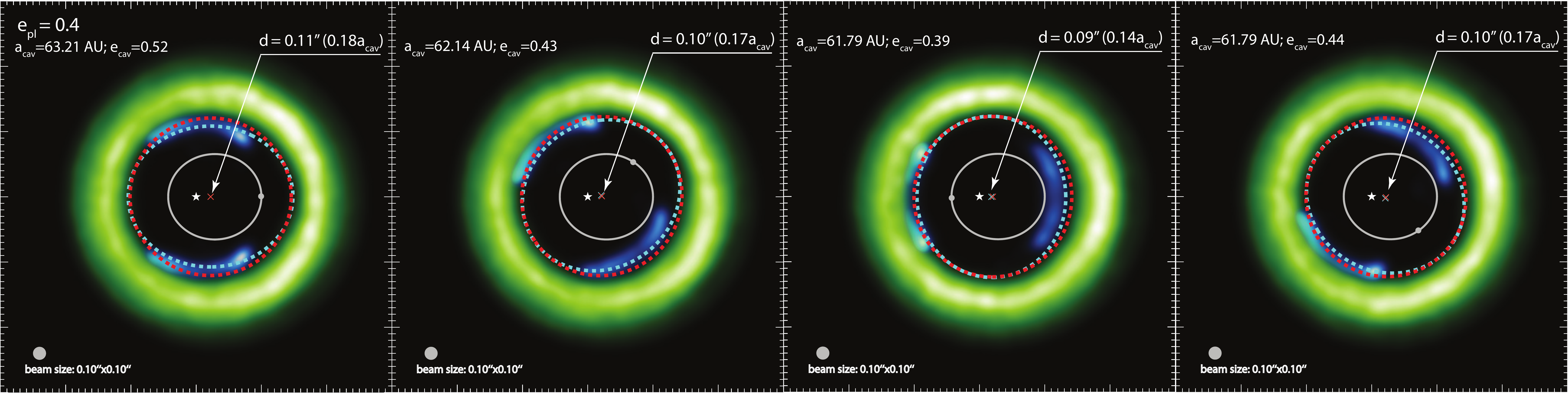}
	\includegraphics[width=2\columnwidth]{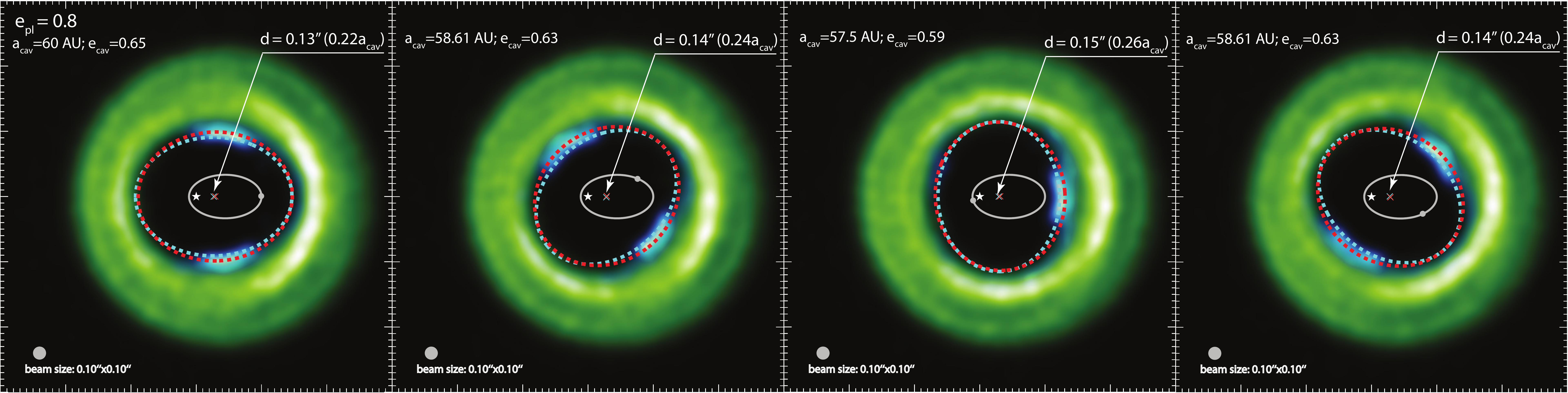}
	\includegraphics[width=2\columnwidth]{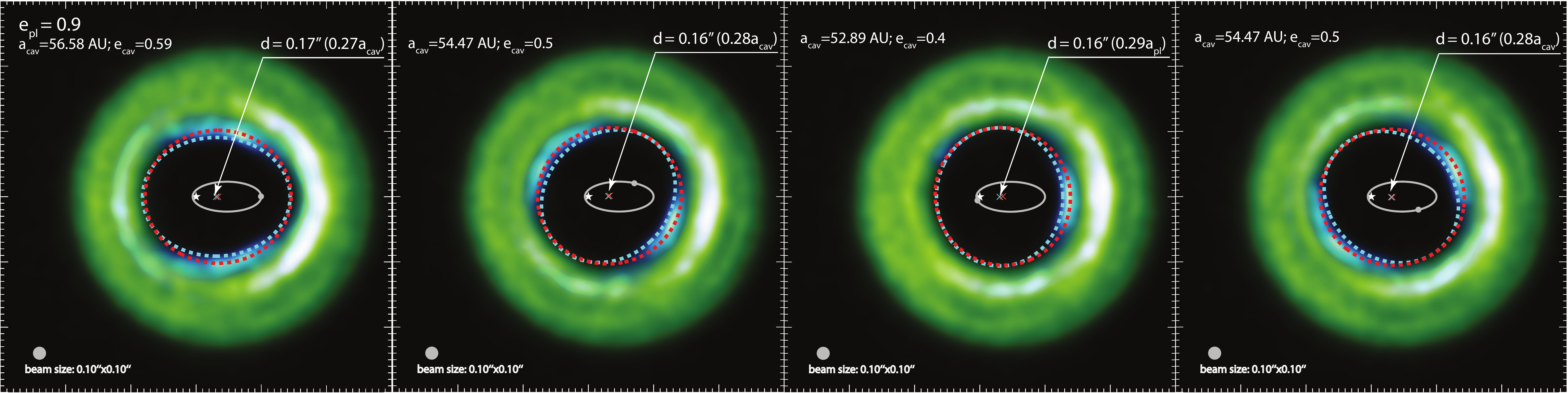}
	\end{center}
	\caption{Change in the cavity edge (dashed blue line) as the $q=5\times10^{-3}$ planet takes a revolution around the star for $e_\mathrm{pl}=0.1,\,0.4,\,0.8$, and $0.9$ models (which represent the four class shown in Fig.\,\ref{fig:a-e}) shown in rows with the parameters of the cavity. The cavity edge of the models without MMRs is also shown (dashed red line). The field of view is 3\arcsec$\times$3\arcsec\,in all panels. Four orbital phases of the planet (true anomaly is $V=\pi,\,5\pi/4,\,0$ and $3\pi/4$) are shown.  The emission of particles in the most populated MMRs (shown with blue and purple colours) are artificially enhanced  to emphasize them (MMRs 1:1 purple and 2:1 blue for $e_\mathrm{pl}=0.1$; MMR 2:1 blue for $e_\mathrm{pl}=0.4$; MMR 3:1 blue for $e_\mathrm{pl}=0.8$, and MMR 4:1 blue for $e_\mathrm{pl}=0.9$). White filled discs and white ellipses represent the planet and its orbit. The white X denotes the centre of the cavity, whose distance measured from the star, $d$ in units of $a_\mathrm{pl}$, is also shown.}
	\label{fig:ch_variation}
    
\end{figure*}

\begin{figure*}
	\begin{center}
	\includegraphics[width=\columnwidth]{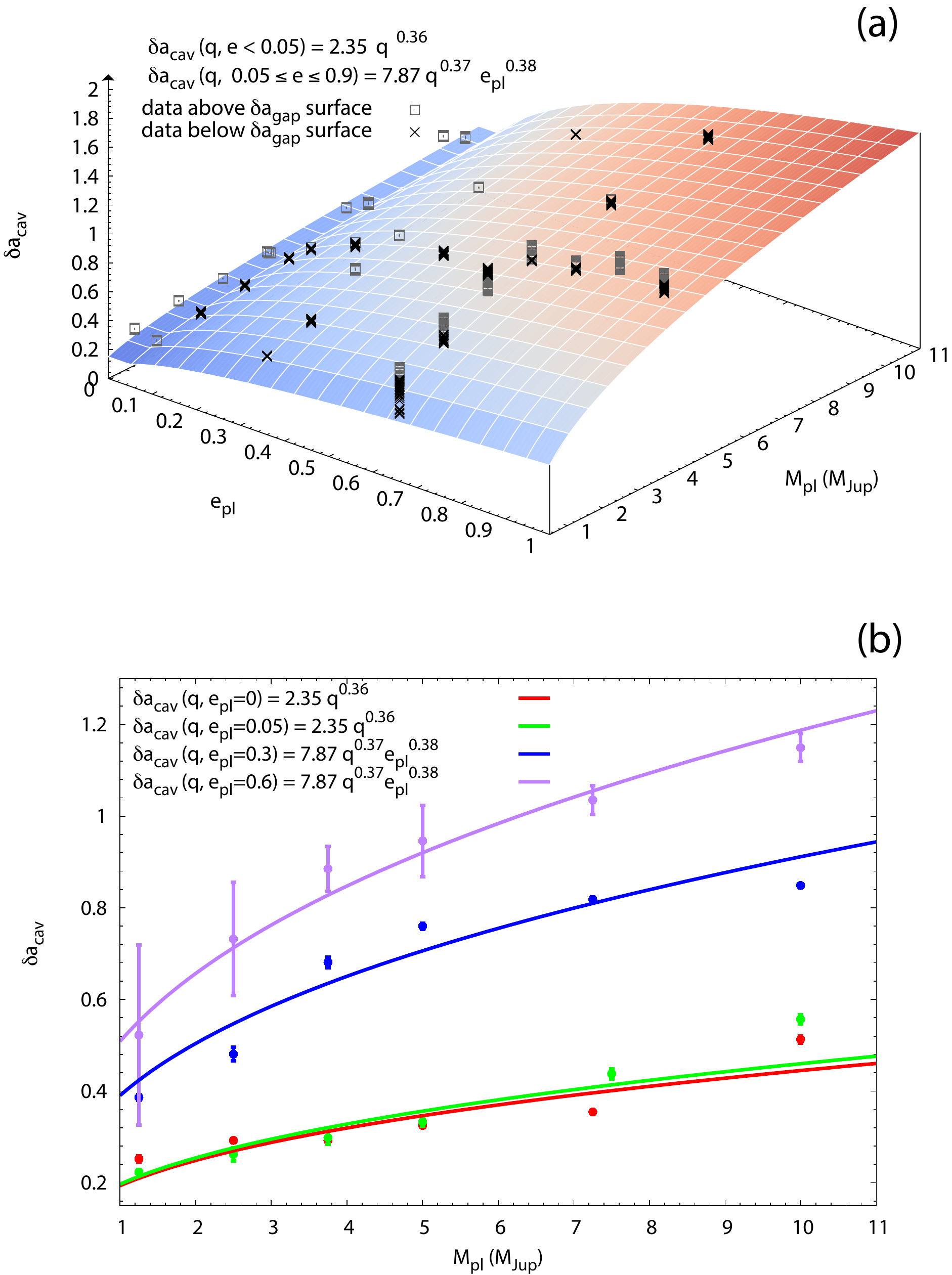}
	\includegraphics[width=0.95\columnwidth]{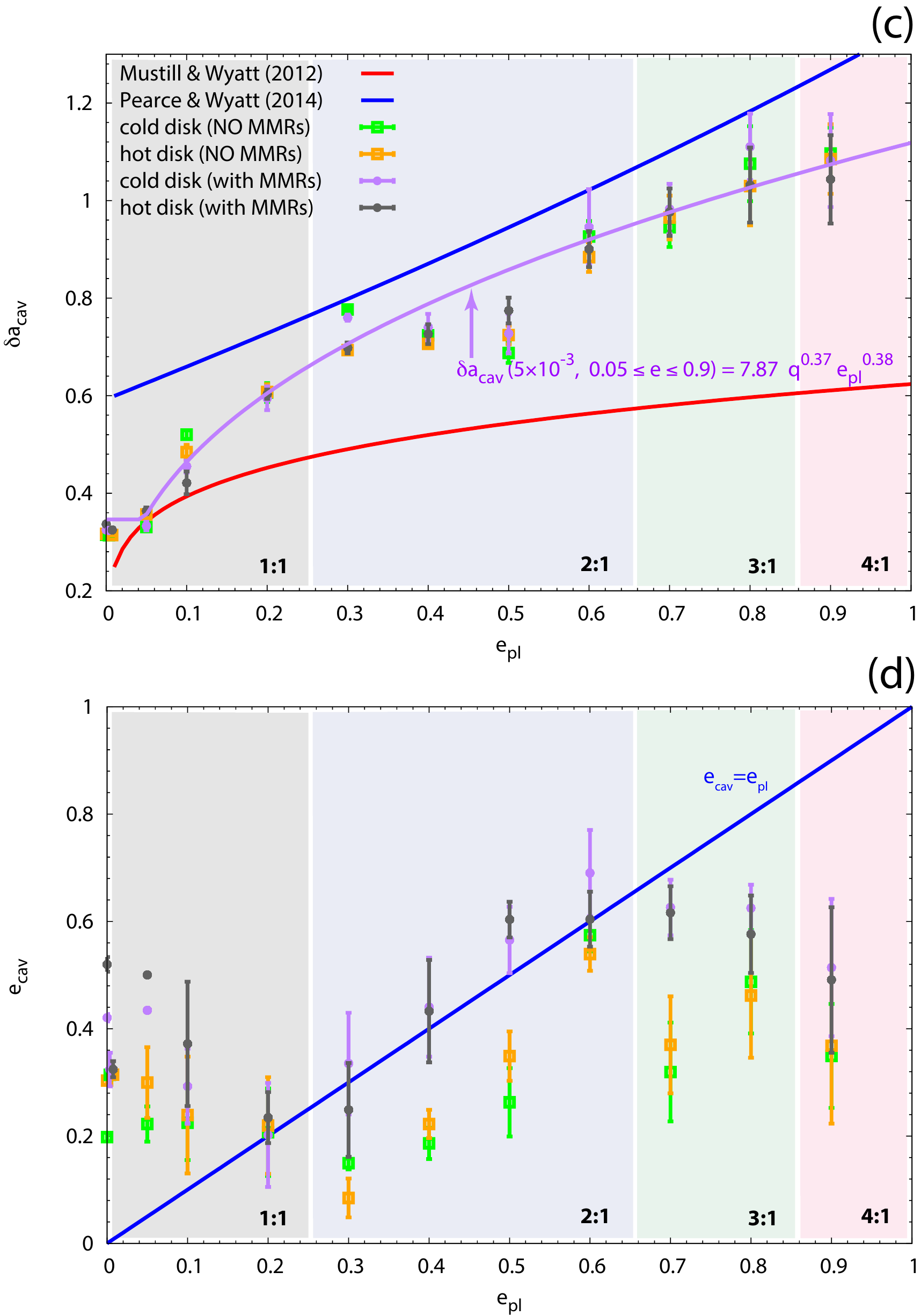}
	\end{center}
	\caption{Panel (a): 2D fit to the cavity size as a function of planetary eccentricity and mass in cold disc models. Panel (b): cavity size versus planetary mass for $e_\mathrm{pl}=0,\,0.05,\,\,0.3$ and $0.6$ in cold disc models. Panels (c) and (d): cavity size and eccentricity versus planetary eccentricity for $q=5\times10^{-3}$ in cold and hot disc models, respectively. $\delta a_\mathrm{cav}$, $e_\mathrm{cav}$ are determined for 40 different orbital phases of the planet. The cavity parameters are shown with purple and grey markers if the emission of MMR particles are taken into account, while orange and green if they are neglected. Former predictions of $\delta a_\mathrm{ch}$  by \citet{MustillWyatt2012} and \citet{PearceWyatt2014} are also shown with red and blue lines, respectively. Error bars represent $2\sigma$ interval of $\delta a_\mathrm{cav}$ and $e_\mathrm{cav}$ during one planetary orbit. Shaded regions emphasize the most populated MMRs in a given models.}
	\label{fig:fit_ch}
\end{figure*}

The orbital elements of particles are determined based on their positions and velocities taken at the end of simulations. Table~\ref{tbl-1} shows the three most populated MMRs for all planetary eccentricities modelled for $q=5\times10^{-3}$ planet. As one can see, four groups can be identified based on that which MMR inside the chaotic zone is the most populated one. Fig.\,\ref{fig:a-e} shows the eccentricity versus semimajor axis of particles for the four different models ($e_\mathrm{pl}=0.1,\,0.4,\,0.8$ and $0.9$), representing the four different groups. The most populated MMR clumps inside the chaotic zone are found to be the 1:1, 2:1, 3:1, and 4:1 for planetary eccentricity in the ranges of $0-0.2$, $0.3-0.7$, $0.8$ and $0.9$, respectively. 3:2 and 5:2 MMRs are populated only for $0\leq e_\mathrm{pl}\leq0.2$ and $0.3\leq e_\mathrm{pl}\leq0.6$, respectively. 2:1 MMR is nearly empty for $e_\mathrm{pl}<0.1$ in agreement with \citep{TabeshianWiegert2016}, while populated for wide range of planetary eccentricities ($0.1<e_\mathrm{pl}\leq0.8$). Populations of  3:1 and 4:1 MMRs can be observed only for $e_\mathrm{pl}\geq0.7$ and $e_\mathrm{pl}\geq0.9$, respectively. Emphasize that all these MMRs, except the 4:1, are inside the chaotic zone (see Fig.\,\ref{fig:a-e}) derived by  \citet{PearceWyatt2014}. 

Fig.\,\ref{fig:ch_variation} shows the thermal emission of the debris discs for the four representative models   ($e_\mathrm{pl}=0.1,\,0.4,\,0.8,$ and $0.9$) for four different orbital phases of the planet.  Fig.\,\ref{fig:ch_variation} also shows the fitted chaotic zone edge by taking into account (ellipse with blue dashed line) and neglecting (ellipse with red dashed line) the emission of MMRs, which are inside the chaotic zone. Note that the emission of particles in the most populated MMRs (shown with blue and purple colours) are artificially enhanced  to emphasize their presence, however, during the chaotic zone fitting this artificial enhancement was not applied. 

The MMR clumps rotate around the star as the planet  takes a full revolution (see reddish and bluish coloured clumps). If significant number of particles are trapped in these MMRs they may affect the size and shape of the planet carved cavity. However, it is appreciable that while the ellipse semimajor axis are nearly independent of the MMR emission, the oblateness of the chaotic zone edge is somewhat smaller if the MMR emissions are neglected. Moreover, independent of the MMR emissions, the cavity ellipse corotates with the planet for $e_\mathrm{pl}\leq0.2$ (top row in Fig.\,\ref{fig:ch_variation}) and rotates with a period which is half that of the planet for $e_\mathrm{pl}\geq0.3$ models (lower three rows of Fig.\,\ref{fig:ch_variation}). This is because of the fact that particles are apsidally aligned with the planet as was shown in \citet{KuchnerHolman2003}.

\subsection{Cavity centre offset}

The cavity always appears offset from the star (the star is not at the geometric centre of the cavity or the disc itself) along the semimajor axis of the planet towards the apocentre (Fig.\,\ref{fig:ch_variation}). The magnitude of the stellar offset increases with the planetary eccentricity. Note that the offset become significant (commensurable to the beam size) for $e_\mathrm{pl}\geq0.2$ models. Empirical formula for the offset distance  (measured in units of the cavity size, $a_\mathrm{cav}$) as a function of the planetary mass and eccentricity can be given in the form $d=c_1q^{c_2}e_\mathrm{pl}^{c_3}$ by applying 2D nonlinear least-squares Marquardt-Levenberg fitting algorithm, which results in
\begin{equation}
	d=0.1q^{-0.17}e_\mathrm{pl}^{0.5},
	\label{eq:doffset}
\end{equation}
with errors in $c_1$, $c_2$, and $c_3$ found to be $\sim5$\%, $\sim6$\%, and $\sim2$\%, respectively. Interestingly, the offset is found to be inversely proportional to the planetary mass. This can be explained by the fact that the 3:4 MMR becomes significantly populated shifting further the fitted ellipse centre from the star for $q\leq1.25\times10^{-3}$, while these resonances are emptied for more massive planets.

The stellar offset is found to be slightly dependent on the planet's orbital phase. As a result, the magnitude of $d$ has an average 10 per cent scatter around its mean value.

\subsection{Brightness asymmetries}
\label{sec:ba}

Another notable property of the disc's millimetre thermal emission is the brightness asymmetry appeared beyond the cavity wall. For a circular planet a weak brightness depression form near the 2:1 MMR (see Fig.\,\ref{fig:calibration}). We note that the clearing of the 2:1 MMR has also been identified by \citet{TabeshianWiegert2016}. Moreover, the azimuthal brightness profile across the peak has $\sim$10 per cent contrast between the minimum and maximum. Note, that this feature is equivalent to the local maximum in the radial brightness profile found by \citet{NesvoldKuchner2015} for their $q\geq10^{-3}$ model. 

For an eccentric planet, the brightness peak always appears close to the apocentre of the planetary orbit (see Fig.\,\ref{fig:ch_variation}). The apocentre glow is permanent independent of the planet's orbital phase. The contrast between the apocentre glow and the fainter parts of the disc increases, while the azimuthal extension of the glow also increases up to $e_\mathrm{pl}=0.2$, then decreases with the planetary eccentricity. By analysing the azimuthal brightness profiles for the investigated planetary eccentricities the followings were found: (1) the brightness difference between the maximum and minimum intensities is found to be in the range of 12  -- 50 per cent; (2) the magnitude of the azimuthal extension of the glow starts with $\sim160\deg$, peaks at $\sim300\deg$, and ends with $\sim120\deg$.
 
The brightening is caused by that the planet forcing particles to eccentric orbits such that the particles apocentres are aligned to the planets apocentre. As a result, an apparent particle concentration appears close to the planet's orbital apocentre, which is dubbed as apocentre glow by \citet{Wyatt2005}. 

By analysing the thermal images we found that the apocentre glow is caused by particles which are trapped in MMRs situated beyond the cavity wall. These MMRs are the 2:1 and 5:2 for $0.1\leq e_\mathrm{pl}\leq0.2$, 3:1 and 4:1 for $0.3\leq e_\mathrm{pl}\leq0.6$, and 4:1 and 5:1 for $0.7\leq e_\mathrm{pl}\leq0.9$.  Although the spatial distributions of resonant particles are found to be slightly varied during a planetary orbit (especially for high planetary eccentricity, $e_\mathrm{pl}\geq0.6$). 

\subsection{Empirical formula for cavity size}
\label{sec:cav_emp}

Both $\delta a_\mathrm{cav}$ and $e_\mathrm{cav}$ are subject to a slight variation during the planetary orbit. Therefore, we calculate orbitally averaged mean values for $\delta a_\mathrm{cav}$ and $e_\mathrm{cav}$ by fitting the cavity for 40 distinct orbital phases of the planet. We found that $\delta a_\mathrm{cav}$ is nearly independent of $e_\mathrm{pl}$ for $e_\mathrm{pl}<0.05$ (panel (a) of Fig.\,\ref{fig:fit_ch}) similarly to what was found by \citet{QuillenFaber2006} for $\delta a_\mathrm{ch}$, however, they derived somewhat larger critical planetary eccentricity of 0.3. \citet{QuillenFaber2006} investigated $q\leq10^{-3}$ regime, therefore we hypothesize that this critical eccentricity decreases with planetary mass. 

Empirical formulas for $\delta a_\mathrm{cav}$ as a function of $q$ and $e_\mathrm{pl}$ is derived by 2D nonlinear least-squares Levenberg-Marquardt algorithm. For  $e_\mathrm{pl}<0.05$ the best-fitting formulae are
\begin{equation}
	\delta a_\mathrm{cav}=2.35q^{0.36}
	\label{eq:da_emp1}
\end{equation}
and for $e_\mathrm{pl}\geq0.05$
\begin{equation}
	\delta a_\mathrm{cav}=7.87q^{0.37}e_\mathrm{pl}^{0.38}.
	\label{eq:da_emp2}
\end{equation}
Note that all the fitted parameters of $\delta a_\mathrm{cav}(e_\mathrm{pl},\,M_\mathrm{pl})$ surface have less than 3\% error.

Panel (b) of Fig.\,\ref{fig:fit_ch} shows $\delta a_\mathrm{cav}$ measured for six different planetary masses ($1.25\times10^{-3}\leq q\leq 10\times10^{-3}$) and empirical formula for the cavity size as a function of $q$ given by Equations (\ref{eq:da_emp1}) and (\ref{eq:da_emp2}) for $e_\mathrm{pl}=0,\,0.05,\,0.3$ and $0.6$. It is notable that the cavity size agrees with the chaotic zone width given by \citet{Wisdom1980} for $e_\mathrm{pl}=0$ and 0.05 models.

Panel (c) of Fig.\,\ref{fig:fit_ch} shows $\delta a_\mathrm{cav}$ measured for different planetary eccentricities ($0\leq e_\mathrm{pl}\leq 0.9$) and our empirical formula for the cavity size given by Equation (\ref{eq:da_emp1}) and (\ref{eq:da_emp2}) for the cold and hot disc models with $q=5\times10^{-3}$. Measurements of $\delta a_\mathrm{cav}$ are shown with and without taking into account the emission of MMR particles. The predictions of the chaotic zone width given by \citet{MustillWyatt2012} ($\delta a_\mathrm{ch}=1.8(q e_\mathrm{f})^{1/5}$) and \citet{PearceWyatt2014} ($\delta a_\mathrm{ch}=5(1+e_\mathrm{pl})[q/(3-e_\mathrm{pl})]^{1/3}$) are also shown. It is appreciable that our method gives values of $\delta a_\mathrm{cav}$ between the former predictions of $\delta a_\mathrm{ch}$. $\delta a_\mathrm{cav}$ is found to be independent of the magnitude of initial eccentricity and inclination of planetesimals as their values are very similar for cold and hot disc models. Note that fluctuations in $\delta a_\mathrm{cav}$ can be observed during a planetary orbit, whose amplitude can reach approximately 10 per cent for high planetary eccentricities ($e_\mathrm{pl}\geq0.5$). This can be explained by that the orbits of particles in populated MMRs are apsidally aligned with the planet such that their apocentre are anti-aligned with that of the planet. As a result, their emissions are highly asymmetric when the planet is at pericentre (see, e.g., the rightmost columns of the two lower rows in Fig.\,\ref{fig:ch_variation}).

Panel (d) of Fig.\,\ref{fig:fit_ch} shows $e_\mathrm{cav}$ versus $e_\mathrm{pl}$ for cold and hot disc models with $q=5\times10^{-3}$. Measurements of $e_\mathrm{cav}$ are shown with and without taking into account the emission of MMR particles.  The mean values of $e_\mathrm{cav}$ are found to be a non-monotonic function of $e_\mathrm{pl}$: its minimum and maximum are at $e_\mathrm{cav}\simeq0.2$ and $e_\mathrm{cav}\simeq0.6$, respectively. $e_\mathrm{cav}$ is also found to be very similar for cold and hot disc models.  We emphasize that  $e_\mathrm{cav}$ and $e_\mathrm{pl}$ agree only in a narrow range of planetary eccentricity, i.e. for $0.3<e_\mathrm{pl}<0.6$.  On the contrary, \citet{PearceWyatt2014} assumed that the chaotic zone edge has the same eccentricity as the planet. Note that the magnitude of the change in $e_\mathrm{cav}$, as the planet takes a full revolution, can be significant (large error bars) for all models.

The  mean values of $\delta a_\mathrm{cav}$ do not change significantly if particles in the MMRs are neglected [see green and orange markers for cold and hot discs, respectively on panel (c) of Fig.\,\ref{fig:fit_ch}]. However, the mean values of $e_\mathrm{cav}$ are significantly lower if particles in the MMRs are neglected [see green and orange markers for cold and hot discs, respectively on panel (d) of Fig.\,\ref{fig:fit_ch}]. Thus, although the particles in MMRs do not affect the semimajor axis of the cavity significantly, the shape of the cavity indeed strongly depends on the emission of those particles. Therefore, the area of the cavity ($a_\mathrm{cav}^2\sqrt{1-e_\mathrm{cav}^2}\pi$), being sensitive to the presence of MMR particles, is larger if we neglect the resonant particles.

\section{Discussion}

\subsection{Comparison to previous works}

For a circular planetary orbit  the chaotic zone width can be given by a simple formula, $\delta a_\mathrm{ch}=1.3q^{2/7}$, based on the theory of overlapping resonances \citep{Wisdom1980}. \citet{QuillenFaber2006} have shown that this formula fits their models assuming $e_\mathrm{pl}\leq0.3$, i.e. $\delta a_\mathrm{ch}$ is independent of the planetary eccentricity. However, \citet{MustillWyatt2012} provided a new formula for the chaotic zone width, $\delta a_\mathrm{ch}=1.8(e_\mathrm{pl}q)^{1/5}$, based on the fact that the width of the MMRs grows with the particle's eccentricity that is significantly pumped up by an eccentric planet. Based on numerical simulations, \citet{PearceWyatt2014} give an approximate expression for the chaotic zone width, $\delta a_\mathrm{ch}\simeq5a_\mathrm{pl,H}=5a_\mathrm{pl}(q/3)^{1/3}$ implicitly assuming that the chaotic zone edge has an elliptic shape whose eccentricity is the same as that of the planet.

In Section\,\ref{sec:cav_emp}, we have shown that the size of the cavity (following an approximate $(qe)^{3/8}$ law profile) is different from that of the chaotic zone given by previous works for an $e_\mathrm{pl}>0.05$ eccentric planet (see Fig.\,\ref{fig:fit_ch}). We have also pointed out that the eccentricity of the cavity edge is always eccentric ($e_\mathrm{cav}\gtrsim0.2$) even for circular planets independent of the emission of MMR particles. Moreover, the cavity edge eccentricity is similar to the planet eccentricity only for $0.3\leq e_\mathrm{pl}\leq0.6$.

In order to explain our results, first, we have to emphasize the difference in the method applied in the previous investigations for determining the cavity size. \citet{QuillenFaber2006} and \citet{Rodigasetal2014} determined the cavity size based on particle lifetime (being equal to the time which is required to gain extremely high eccentric orbits) measurements. \citet{MustillWyatt2012} select chaotic orbits by the high number of peaks in the Fourier transform of the eccentricity evolution spectrum. Another method to measure the width of the cavity is based on calculating the half-brightness distance of the radial brightness profile of the debris disc \citep{Chiangetal2009,NesvoldKuchner2015}. These studies implicitly assumed that the cavity edge is circular or similar to that of the planet. The radial brightness profile is applicable as long as the cavity edge is circular, which is not the case as we have presented in Fig.~\ref{fig:fit_ch}. This is why we fit an ellipse to the the thermal emission of the cavity edge rather than measuring the radial brightness profile, which resulted in somewhat smaller $\delta a_\mathrm{cav}$ than that of \citet{MustillWyatt2012}.

Further notable differences to previous numerical investigations of the chaotic zone is that former studies used an order of magnitude smaller number of test particles. This can cause high Poisson noise of synthetic images or low signal-to-noise ratio of radial brightness profiles. To mitigate the Poisson noise one can average together several (10-50) outputs of integration steps (see, e.g., \citealp{PearceWyatt2014,NesvoldKuchner2015}).  Another possibility is to spread each surviving particles out along its orbit, in which case, a particle is cloned and placed at discrete locations along its orbit (see, e.g. \citealp{Chiangetal2009,Rodigasetal2014}). However, these procedures may affect the inferred size of the cavity (presumably increases) due to the fact that the cavity takes a full revolution during a planetary orbit. We emphasize that we do not use any of these procedures, as in our simulations there are quarter of million survived particles in the disc at the converged equilibrium state. Thanks to our synthetic image calculation method, we were able to identify azimuthal brightness asymmetries in the debris disc similarly to an additional peak in radial brightness profile developed beyond 2:1 MMR found by \citet{NesvoldKuchner2015}.

Here we have to mention some caveats of our models. We applied the restricted three-body approach, thus the perturbation on the planetary orbit such as planetary precession \citep{PearceWyatt2015} or migration and subsequent fail of trapping particles in MMRs for an eccentric planetary orbit ($e_\mathrm{pl}>0.075$, \citealp{MustillWyatt2011,Recheetal2008} are neglected. However, our assumption is plausible as long as the mass of the disc is much less than that of the giant planet.  

In our investigation, we did not assume further planets orbiting in the system. Additional planet(s) exchanging significant amount of angular momentum with the sculpting planet can cause fluctuation in its eccentricity. In this case, a high eccentricity of the sculpting planet could suggest the existence of a second planet in the system.

We applied a collisionless model in which case the spatial distribution of millimetre sized dust particles follow that of the planetesimals. \citet{StarkKuchner2009} have shown that collisional destruction of grains is enhanced in MMR. Considering collisions of planetesimals assuming an embedded several Jupiter mass planet on circular orbit, \citet{NesvoldKuchner2015} found that MMRs are depleted in planetesimals (except for the 1:1 MMR) resulting in $15-25$ per cent larger cavity size (see their fig.~2) than the previous prediction of \citet{MustillWyatt2012}. Note that our study also gives larger cavity sizes by about $15$ per cent than that of \citet{MustillWyatt2012}. Nevertheless, to investigate the effect of MMR planetesimals on the cavity size, we artificially removed their emission. We found that MMR particles have no significant effect on $\delta a_\mathrm{cav}$ (Fig.~\ref{fig:fit_ch}), however, their absence results in smaller but non-zero $e_\mathrm{cav}$. Thus, we think that although taking into account the collisions of planetesimals can be essential, the cavity carved by the giant planet is not identical to the chaotic zone because of  the cavity wall is not circular.

Note that care must be taken applying the presented method on scattered light images as the radiation pressure (possibly smearing out resonant structures) can not be neglected for small grains which dominate those images.  Finally, we did not consider multiple planets orbiting in the system for which case our empirical formula might also be inadequate.

\subsection{Synthetic ALMA images}
\label{sec:observability}

To investigate observable properties of the cavity carved by a giant planet synthetic ALMA images are calculated in the millimetre wavelengths with ALMA Observation Support Tool version 5.0 for four $q=5\times10^{-3}$ models assuming $e_\mathrm{pl}=0.05,\,0.1,\, 0.4$ and $0.8$. The source model for the synthetic ALMA images is the thermal emission of simulated discs (see Fig.~\ref{fig:ch_variation}) assuming source distance and size of $100$\,pc (top row of Fig.~\ref{fig:ALMA}) and $a_\mathrm{pl}=50$\,au (lower row of Fig.~\ref{fig:ALMA}), respectively. \citet{NesvoldKuchner2015}found that MMRs could be unpopulated due to effect of collisions. To consider this effect the emission of particles trapped in MMRs inside the chaotic zone (MMR\#1 and \#2 presented in Table~\ref{tbl-1}) are removed. The source position on the sky resembles that of HD\,95086, i.e., $\delta$=-68$^{\circ}$40\arcmin2.5\arcsec. The synthetic images are obtained for C$43-5$ antenna configuration resulting in $ \bar{\sigma}\simeq0.37\arcsec$ beam size ($\bar{\sigma}=(\sigma_\mathrm{x}+\sigma_\mathrm{y})/2$). The central frequency and bandwidth of ALMA observations are 230.16\,GHz and 7.5\,GHz, respectively.  We assume an average atmospheric condition with  precipitable water vapour of $1.796$\,mm. The resulting images are deconvolved with the {\sc CLEAN} algorithm using natural weighting.  The on-source time is set to 3~h, assuming optimal zenith distance, i.e 1.5~h exposure prior and after culmination of the target.  The surface density of the discs, i.e the model image brightness is scaled such that the synthetic images have a total flux of about $2$\,mJy, which corresponds to the flux densities scaled to 100\,pc of the brightest known debris discs at 1.3\,mm (see, e.g., \citealp{Marinoetal2016,Lieman-Sifryetal2016}).

\subsection{Measurements on ALMA images}
\label{sect:ALMAmeas}

\begin{figure}
	\begin{center}
	\includegraphics[width=\columnwidth]{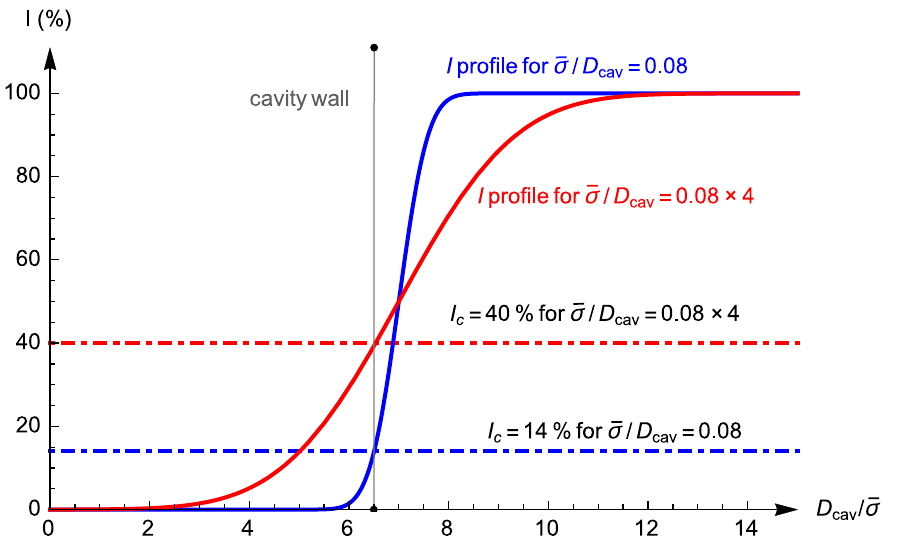}
	\end{center}
	\caption{Effect of the image resolution on the brightness profiles (solid lines) across the cavity wall: $D_\mathrm{cav}/\bar{\sigma}\simeq12.5$ (blue) and $12.5/4$ (red) obtained for the calibration images and synthetic ALMA images, respectively. The critical intensity levels for the calibration images (dot-dashed blue) and the synthetic ALMA images (dot-dashed red) are also shown.}
	\label{fig:effres}
\end{figure}

\begin{figure*}
	\begin{center}
	\includegraphics[width=2\columnwidth]{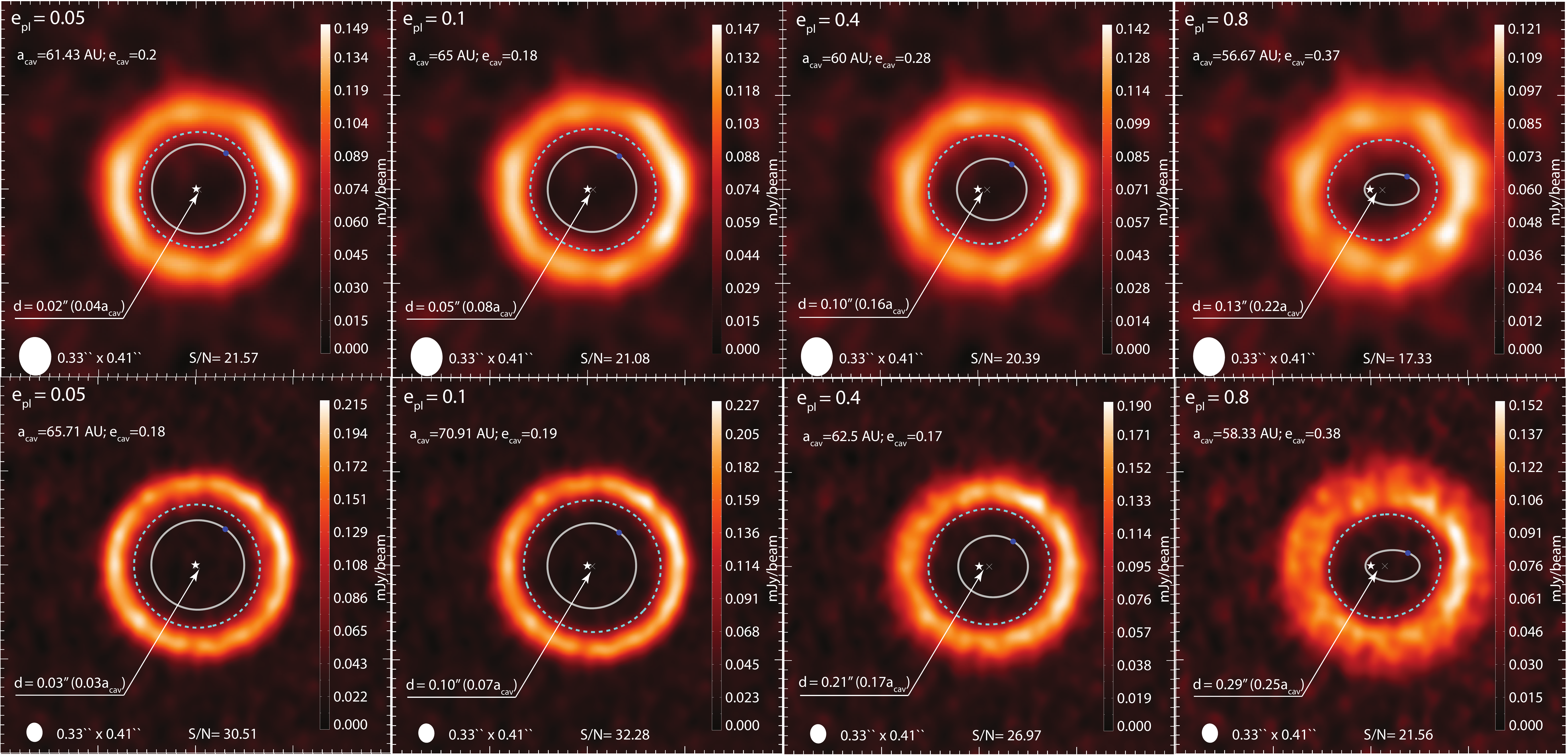}
	\end{center}
	\caption{Synthetic ALMA images (assuming C43-5 antenna configuration) calculated for $q=5\times10^{-3}$ and $e_\mathrm{pl}=0.05, 0.1, 0.4 $ and $ 0.8$ models. The source distance is assumed to be 100\,pc (top panel) and 50\,pc (bottom panel). The emission of the most populated MMR~1 and MMR~2  (which are detached from the disc, see Table~\ref{tbl-1}) are removed from the source models. The field of view is $4\arcsec\times4\arcsec$ and $8\arcsec\times8\arcsec$ for the top and bottom panels, respectively. The dashed and solid lines represent the fitted cavity edge and planetary orbit, respectively. The blue filled circles show the planet's orbital position. The obtained signal-to-noise ratios (defined as the ratio of the peak intensity to the rms on the image where there is no source emission) are also shown. The offsets of the stellar position and the cavity centre, $d$, are also shown.}
	\label{fig:ALMA}
\end{figure*}

To measure the cavity properties seen on synthetic ALMA images first an appropriate intensity level, $I_\mathrm{c}$, is required as described in Section\,\ref{sec:dust-em}. The appropriate intensity level for the maximum normalized thermal images having $\sigma=0.1\arcsec$ is found to be $I_\mathrm{c}^0=14\%$. However, $I_\mathrm{c}$ depends on the image resolution.  The slope of the brightness profile at the cavity wall decreases with increasing beam size, therefore the image brightness is somewhat higher at the defined position of the cavity edge for lower image resolution.  The source distance has the same effect, i.e. if the source is resolved by smaller number of beams, higher value of $I_\mathrm{c}$ is necessary, see Fig.~\ref{fig:effres}. 

We define the effective image resolution expressed by the number of beams resolving the cavity, $D_\mathrm{cav}/\bar{\sigma}$, where $D_\mathrm{cav}$ is the estimated diameter of the cavity in arc seconds. The brightness profile of the cavity wall can be approximated by the convolution of a perfect step function (the cavity wall seen with infinitely large resolution) and a Gaussian function assuming full width at half maximum being equal to the effective beam size of the ALMA observation, which reads
\begin{equation}
I=\frac{1}{2}\left(1+\mathrm{Erf}\left(\frac{x_\mathrm{cav}}{\bar{\sigma}/D_\mathrm{cav}}\right)\right),
\label{eq:ic}
\end{equation}
where
\begin{equation}
\mathrm{Erf}(x)=\frac{2}{\sqrt{\pi}}\int^x_{0}e^{-z^2}dz.
\end{equation}
Assuming that the critical intensity level is $I_\mathrm{c}=0.14$ and $D_\mathrm{cav}/\bar{\sigma}\simeq12.5$  (see our calibration process on Fig.~\ref{fig:calibration}), the numerical solution of the Equation~(\ref{eq:ic}) gives  $x_\mathrm{cav}\simeq-0.06$. The cavity wall on synthetic ALMA images having an effective resolution of $D_\mathrm{cav}/\bar{\sigma}\simeq12.5/4$ (top row of Fig.~\ref{fig:ALMA}) and $D_\mathrm{cav}/\bar{\sigma}\simeq12.5/2$ (lower row of Fig.~\ref{fig:ALMA}) should be at the same position as in the calibration image, therefore the critical intensity levels set to $I_\mathrm{c}\simeq0.4$ and $I_\mathrm{c}\simeq0.3$, respectively.

To determine the orbital parameters of a putative planet that carves the cavity observed on continuum ALMA images, we propose the following method: (1) Deproject the image (assuming that the outer edge of the debris disc has a circular shape\footnote{By analysing the discs' global shape in $q=5\times10^{-3}$ models assuming different planetary eccentricity we found that the disc edges are nearly circular ($e_\mathrm{dic}\lesssim0.1$) for all cases.}) with the known source inclination and normalize the image with the maximum pixel intensity;  (2) estimate the effective image resolution, $D_\mathrm{cav}/\bar{\sigma}$, on the deprojected image; (3) calculate the appropriate contour level, $I_\mathrm{c}$, for the cavity edge fitting by using the following expression  
\begin{equation}
I_\mathrm{c}\simeq\frac{1}{2}\left(1+\mathrm{Erf}\left(\frac{-0.06}{\bar{\sigma}/D_\mathrm{cav}}\right)\right);
\label{eq:critInt}
\end{equation}
(4) to infer the contours of the cavity edge, select pixels on the image which have intensities close to $I_\mathrm{c}$; (5) fit an ellipse to the selected contours and determine the cavity size, $a_\mathrm{cav}$; (6) determine the stellar offset, $d$, measured from the cavity centre.

By measuring the cavity size one can infer the orbital parameters of a putative planet that carved the cavity if the planet mass is known. However, as $\delta a_\mathrm{cav}$ is degenerated [see, e.g., Equation\,(\ref{eq:da_emp2}] which includes both $q$ and $e_\mathrm{pl}$), only a combined range of $a_\mathrm{pl}$ and $e_\mathrm{pl}$ can be given. Thus, for a given star-to-planet mass ratio, $q$, and $e_\mathrm{pl}$ the cavity size can be used to infer a plausible value for $a_\mathrm{pl}$ as
\begin{equation}
	a_\mathrm{pl}=\frac{a_\mathrm{cav}}{1+7.87q^{0.37}e_\mathrm{pl}^{0.38}},
	\label{eq:a_pl-a_cav}
\end{equation}
where we use the definition of the cavity size, $a_\mathrm{cav}=(1+\delta a_\mathrm{cav})a_\mathrm{pl}$, and the empirical relation for $\delta a_\mathrm{cav}$ given by Equation\,(\ref{eq:da_emp2}). Nevertheless, if a stellar offset ($d$ measured in the unit of cavity size) is appreciable on the continuum image, $e_\mathrm{pl}$ can be estimated as
\begin{equation}
	e_\mathrm{pl}=\left(\frac{d}{0.1q^{-0.17}}\right)^{2},
	\label{eq:e_pl-d}
\end{equation}
where we use the empirical relation for the stellar offset given by Equation\,(\ref{eq:doffset}). Note that both formulae have intrinsic uncertainties, which are discussed in the next section.

\subsection{Observability with ALMA}

We determined the cavity sizes and eccentricities for $q=5\times10^{-3}$ and $0\leq e_\mathrm{pl}\leq0.9$ models from the synthetic ALMA images (four particular cases are shown in Fig.\,\ref{fig:ALMA} assuming 50 and 100\,pc source distances). The average error in $a_\mathrm{cav}$, are found to be modest, 6 per cent and 10 per cent (independent of planetary eccentricity) for 50 and 100\,pc source distances, respectively. However, the error in the cavity eccentricities are somewhat larger, 15 per cent and 32 per cent for the same source distances.

We have shown that the offset of the stellar position with respect to the cavity centre becomes significant for $e_\mathrm{pl}\gtrsim0.2$ (see Section~4.3). These offsets are significant on the synthetic ALMA images, see $e_\mathrm{pl}\geq0.4$ models on Fig.\,\ref{fig:ALMA}. Detecting such offset might be an indication for an eccentric massive planet, if the pointing error of ALMA antenna system is less than the observed offset.

As we presented in Section~\ref{sec:ba} an apocentre glow develops for eccentric planets whose brightness increases with $e_\mathrm{pl}$. To determine the significance of the apocentre glow we calculated the disc's azimuthal brightness profiles, on which the contrast as the difference between the maximum and mean flux is measured. We found that the apocentre glow has a $\sim5\sigma_{\mathrm{noise}}$ significance, where $\sigma_{\mathrm{noise}}$ is the standard deviation of the naturally weighted synthetic ALMA image. 

The azimuthal brightness profiles of the discs are double-peaked for $e_\mathrm{pl}<0.1$ models with $\bar{\sigma}/D_{\mathrm{cav}}=12.5/4$ effective resolution (as one can see on the first two top panels of Fig.~\ref{fig:ALMA}) due to the elliptical beam shape caused by the assumed declination of our synthetic target. However, the profiles are single-peaked (as it should be explained in Section~\ref{sec:ba}), if the beam shape is circular. If the source is at 50\,pc distance for which case the effective resolution is doubled we do not get false glow. For $e_\mathrm{pl}>0.1$ do not appear artificial glows. To summary, care must be taken to interpret brightness asymmetries on ALMA images if the effective resolution is lower than about $3$, because in this case an artificial glow can appear on the pericentre side of the disc causing a double-peaked brightness profile.

We have also shown that the cavity wall rotates as the planet orbits the star (with a period equals to that of the planet period for $e_\mathrm{pl}\leq0.2$ and half of that if $e_\mathrm{pl}\geq0.3$). By analysing the synthetic ALMA images, we found that the rotation of the cavity edge can not be observed if the effective resolution is $D_\mathrm{cav}/\bar{\sigma}\simeq3$ (see top panel of Fig.~\ref{fig:ALMA}), being significantly smaller than $D_\mathrm{cav}/\bar{\sigma}=12.5$ for which case the rotation is visible, see Fig.\,\ref{fig:ch_variation}. The cavity edge rotation would be observable with larger ALMA baselines providing better resolution. However, with better angular resolution the achievable signal-to-noise ratio is not high enough ($SN\leq10$) to sufficiently fit the cavity edge, assuming a plausible source brightness $\sim2$\,mJy total flux for the source at 100\,pc.  Nevertheless, for nearby sources, e.g. at a distance of $50$\,pc, the cavity rotation becomes visible on the synthetic ALMA images (see bottom panel of Fig.\,\ref{fig:ALMA}). This can be explained by that the effective resolution doubles ($D_\mathrm{cav}/\bar{\sigma}\simeq6$), while the source brightness quadruples ($8$\,mJy). If the source brightness were smaller (e.g., $4$\,mJy at 50\,pc), the cavity rotation would still be visible.

\subsection{Estimations from ALMA observations}

To test our method proposed for determining the orbital parameters of a putative planet that carves the observed cavity in a debris disc, we investigated three scenarios: (I) no planet is known in the system; (II) planet location is known, but its mass is completely unknown; and (III) both planet location and mass are known with a certain accuracy.

In scenario I, by measuring $a_\mathrm{cav}$ one can infer plausible $q-a_\mathrm{pl}$ pairs for a given $e_\mathrm{pl}$. By measuring $d$, the plausible $e_\mathrm{pl}$ can be constrained by Equation\,(\ref{eq:e_pl-d}). Assuming that the planet has a low eccentric orbit, $e_\mathrm{pl}\leq0.05$, one can constrain the planetary semimajor axis. Since it is reasonable to assume that $q\leq1$ Equation\,(\ref{eq:da_emp1}) leads us to a constraint for the planetary semimajor axis of $a_\mathrm{pl}\gtrsim 0.3a_\mathrm{cav}$, and obviously $a_\mathrm{pl}<1a_\mathrm{cav}$. 

\begin{figure}
	\begin{center}
    \includegraphics[width=\columnwidth]{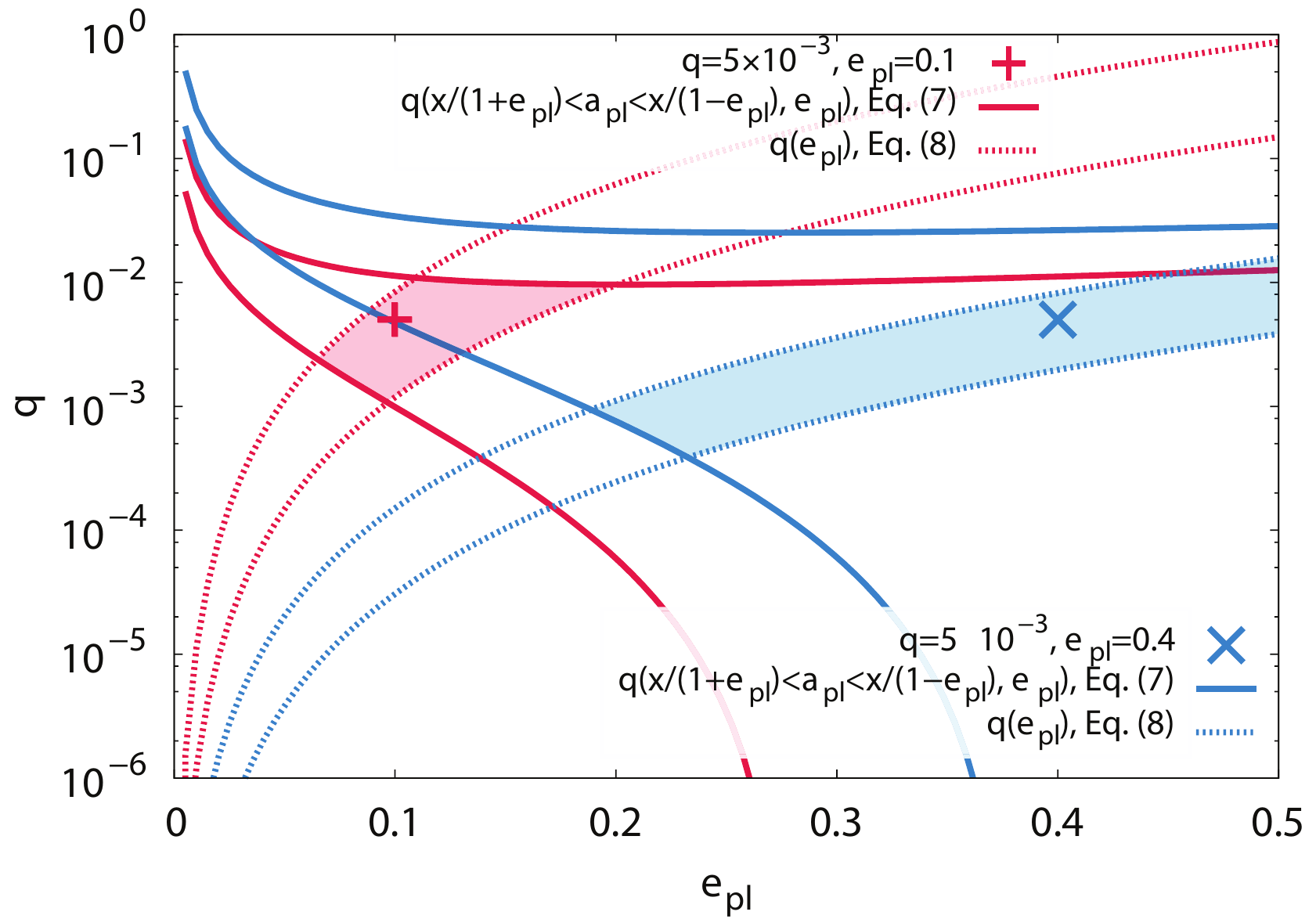}
	\includegraphics[width=\columnwidth]{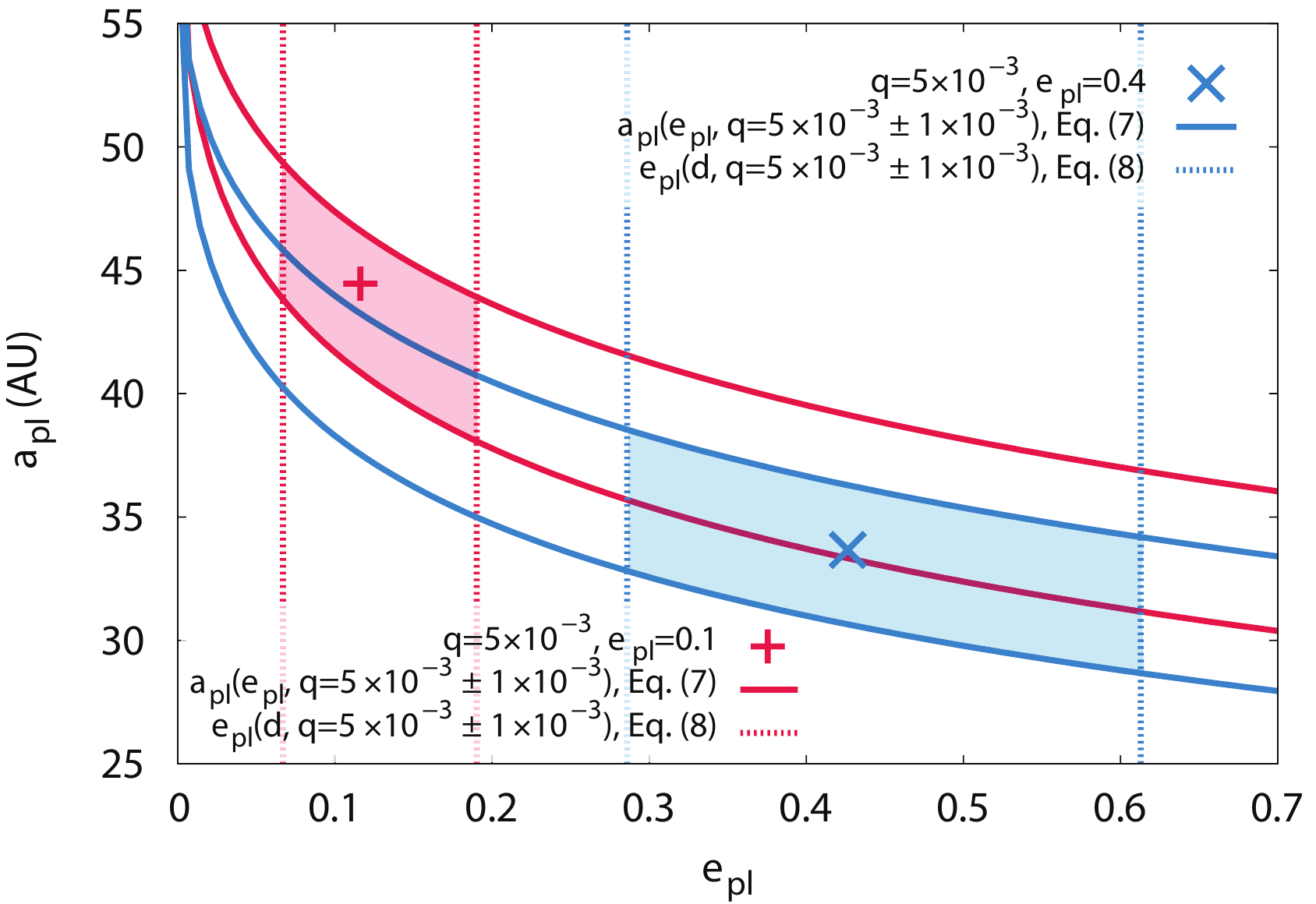}
	\end{center}
    \caption{Top panel: estimating $q$ and $e_\mathrm{pl}$ with the assumption of that the planet mass is unknown (scenario II). Bottom panel: estimating $a_\mathrm{pl}$ and $e_\mathrm{pl}$ with the assumption of that the planet mass is known with 20 per cent accuracy (scenario III). The measurements are done on models $q=5\times10^{-3}$ and $e_\mathrm{pl}=0.1$ and $0.4$ (represented by + and $\times$) using $a_\mathrm{cav}$ and $d$ taken from synthetic ALMA images shown in the top row of Fig.\,\ref{fig:ALMA}. Solid and dashed lines show the possible ranges, while the shaded regions are the most probable values of $q$, $a_\mathrm{pl}$ and $e_\mathrm{pl}$.}
	\label{fig:fit_test}
\end{figure}

In scenario II, by measuring $a_\mathrm{cav}$ and $d$ one can infer the most probable values for $q$ and $e_\mathrm{pl}$ if the planet location, $x_\mathrm{pl}$, is known. We measured cavity parameters and the planet-to-star distance for models $q=5\times10^{-3}$, $e_\mathrm{pl}=0.1$ and $0.4$ on our synthetic ALMA images assuming 100\,pc source distance (top row of Fig\,\ref{fig:ALMA}). Since $(1-e_\mathrm{pl})a_\mathrm{pl}\leq x_\mathrm{pl}\leq(1+e_\mathrm{pl})a_\mathrm{pl}$ Equation\,(\ref{eq:a_pl-a_cav}) leads us to a constraint for $q$ shown by solid curves on the top panel of Fig.\,\ref{fig:fit_test}. Moreover, Equation\,(\ref{eq:e_pl-d}) also leads us a constraint for a plausible range of $e_\mathrm{pl}$ shown by dotted curves on the top panel of Fig.\,\ref{fig:fit_test}. As the fitted parameters of the empirical relations Equations\,(\ref{eq:a_pl-a_cav}) and (\ref{eq:e_pl-d}) have intrinsic errors, while $a_\mathrm{cav}$ and $d$ have $\sim10$ per cent uncertainties (see Sections\,4.3 and 4.5), two-two curves are plotted for the plausible $q$ and $e_\mathrm{pl}$. Thus, the most probable values of $q$ and $e_\mathrm{pl}$ are bounded by the solid and dashed curves. Taking into account both most probable values of $q$ and $e_\mathrm{pl}$ one can further constrain those parameters shown by the shaded regions on the top panel of Fig.\ref{fig:fit_test}. The original model parameters marked by + and $\times$ lie inside the shaded regions. As a result, by measuring cavity parameters and planet position, one can infer $q$ with an order of magnitude uncertainty.

In scenario III, by measuring $a_\mathrm{cav}$ and $d$ one can infer $a_\mathrm{pl}$ and $e_\mathrm{pl}$ by using Equations\,(\ref{eq:a_pl-a_cav}) and (\ref{eq:e_pl-d}), if $q$ is known with a certain accuracy. We tested our method on the same models that previously were used with an assumption of that $q=5\times10^{-3}$ with 20 per cent accuracy (like for HD\,95086b, see, e.g., \citealp{Rameauetal2013}). The results are shown on the bottom panel of Fig.\,\ref{fig:fit_test}. Here we have taken into account the cumulative errors in the fitted parameters of Equations\,(\ref{eq:a_pl-a_cav}) and (\ref{eq:e_pl-d}) and the $\sim10$ per cent uncertainty of $a_\mathrm{cav}$ and $d$ discussed in Sections\,4.5 and 4.3, respectively. As one can see, $a_\mathrm{pl}$ can only be determined with a $\sim10$ per cent error (shown by regions bounded by solid curves). However, if the planetary orbit is nearly circular (e.g., $e_\mathrm{pl}\lesssim0.1$), the uncertainty of $a_\mathrm{pl}$ decreases to $\sim4$ per cent. $e_\mathrm{pl}$ can also be estimated by an appreciable stellar offset with a relatively large error of $\sim20$ per cent and $\sim15$ per cent (shown by dashed vertical lines) for models $e_\mathrm{pl}=0.1$ and $0.4$, respectively.

\section{Summary and Conclusion}

In this paper, we investigated the clearing of the chaotic zone of a giant planet embedded in a debris disc. \emph{N}-body simulations were performed with a GPU-based high-precision Hermite direct integrator in 3D. We assumed that the debris material consists of collisionless  bodies whose dynamics are affected only by the central star and gravitational perturbation of a $1.25-10\,M_\mathrm{Jup}$ giant planet. We modelled dynamically cold and hot discs assuming planetary eccentricity ($e_\mathrm{pl}$) in the range of $0-0.9$. The outer edge of the cavity was determined by fitting an ellipse to the millimetre wavelength ($\lambda=1.3$\,mm) thermal emission of the dust. The cavity edge fitting method is calibrated such that the cavity size agrees with the previous prediction of  chaotic zone size given by \citet{Wisdom1980} for $e_\mathrm{pl}=0$. 

Based on our \emph{N}-body simulations, we conclude that the chaotic zone and the cavity observed in debris discs are not identical, in general. Therefore, our new empirical formulae (Equations \ref{eq:da_emp1} and \ref{eq:da_emp2}) for the size of a giant planet carved cavity gives a more realistic values than previous estimations, especially for an eccentric giant planet.

Synthetic ALMA images were also calculated with a resolution of $\bar\sigma=0.37\arcsec$ mean beam size provided by the C$43-5$ antenna configuration assuming $\simeq2$\,mJy total flux emitted by the disc at millimetre wavelengths. To take into account the MMR removal by collisions \citep{NesvoldKuchner2015}, the most populated MMRs were removed from the source models of synthetic ALMA images. A new method for the determination of the planetary orbital parameters based on high-resolution ALMA observations (the cavity should be resolved by more than about 20 beams) are also given. Our main findings are the followings:

\begin{enumerate}
\item{Independent of the planetary eccentricity, the cavity wall is not circular but has an elliptic shape. The cavity edge corotates with the planet with a period equals to and half that of the planet for $e_\mathrm{pl}\leq0.2$ and $e_\mathrm{pl}\geq0.3$, respectively (Fig.\,\ref{fig:ch_variation}). }
\item{The cavity centre is off-centred along the semimajor axis of the planet towards the apocentre. The magnitude of the offset is found to be $d\simeq0.2e_\mathrm{pl}^{0.8}a_\mathrm{pl}^{-0.2}$ (Fig.\,\ref{fig:ch_variation}).}
\item{An apocentre glow develops for $e_\mathrm{pl}>0$ with 12 per cent $-$ 50 per cent (increasing with $e_\mathrm{pl}$) contrast between the bright and faint parts of the disc. The azimuthal extension of the glow  $\gtrsim120\deg$, and peaks at $\sim300\deg$ for $e_\mathrm{pl}=0.2$.}
\item{The cavity size represented by the semimajor axis of a fitted ellipse to cavity edge is found to be $\delta a_\mathrm{cav}\simeq2.35q^{0.36}$ for $e_\mathrm{pl}<0.05$ and $\delta a_\mathrm{cav}\simeq7.87q^{0.37}e_\mathrm{pl}^{0.38}$ for $e_\mathrm{pl}\geq0.05$. Our empirical formula gives $\delta a_\mathrm{cav}$ values between the chaotic zone sizes given by \citet{MustillWyatt2012} and  \citet{PearceWyatt2014}, independent of the dynamic temperature of the disc and the emission of particles trapped in MMRs inside the chaotic zone (Fig.\,\ref{fig:fit_ch}).}
\item{The eccentricity of the cavity edge, $e_\mathrm{cav}$, is found to be a non-monotonic function of $e_\mathrm{pl}$: $e_\mathrm{cav}$ is larger for $e_\mathrm{pl}\leq0.2$, nearly equals for $0.3\leq e_\mathrm{pl}\leq 0.6$, and smaller for $e_\mathrm{pl}\geq0.7$ than the planetary eccentricity. $e_\mathrm{cav}$ is significantly smaller if the emission of MMR particles are neglected (Fig.\,\ref{fig:fit_ch}).}
\item{Our proposed method for determining the cavity size with ALMA observations, which takes into account the effective resolution ($D_\mathrm{cav}/\bar{\sigma}$, number of beams resolving the cavity) gives appropriate values within a 10 per cent error. }
\item{Assuming a source at 100\,pc with a total flux of $2$\,mJy,  the apocentre glow is detectable with a significance level of 5. However, if the cavity is resolved by less than about three beams having an elliptical shape artificial brightness peaks may appear opposite to the apocentre for $e_\mathrm{pl}<0.1$. With this effective resolution the cavity rotation can not be detected on synthetic images. In contrast, if the source is at 50\,pc (the cavity is resolved by about six beams) the cavity rotation is detectable even for a 4\,mJy total source flux.} 
\end{enumerate}

Measurements of the size of the planet carved cavity and stellar offset on high-resolution millimetre observations with ALMA can only predict plausible ranges for $q$ and orbital parameters ($a_\mathrm{pl}$ and $e_\mathrm{pl}$) if $q$ is known with a certain accuracy. This is due to the fact that $\delta a_\mathrm{pl}$ depends on both $q$ and $e_\mathrm{pl}$. 

Our numerical experiments done on synthetic ALMA images showed that $a_\mathrm{pl}$ and $e_\mathrm{pl}$ can be determined with $\lesssim10$ per cent and $\lesssim20$ per cent accuracies if the planet-to-star mass ratio is known with $20$ per cent error [see bottom panel of Fig.\,\ref{fig:fit_test}]. If the planet mass is unknown $q$ can only be inferred with an order of magnitude accuracy (see top panel of Fig.\,\ref{fig:fit_test}).

We emphasize that the planetary eccentricity can not be inferred by measuring the cavity eccentricity due to the degeneracy of $e_\mathrm{cav}-e_\mathrm{pl}$ relation (see panel (d) of Fig.\,\ref{fig:ch_variation}). However, apocentre glow with high contrast favours orbital solutions with significant planetary eccentricity ($e_\mathrm{pl}\gtrsim0.1$). Note, however, that care must be taken when analysing ALMA observations with highly elliptical shape beam, which can produce false asymmetries. Moreover, care also must be taken for highly inclined debris discs ($i>50^\circ$) as brightness peaks can be an artefact of image reconstruction \citep{Millietal2012}.

Due to the serious resolution requirement, our empirical formulae are best applicable for nearby bright debris discs, such as Fomalhaut, HR\,4796, HD\,202628, HD\,181327, HD\,107146,\, and HD\,95086.  Scattered light images of some of these discs are consistent with an eccentric cavity: e.g., $e_\mathrm{cav}=0.11$ for Fomalhaut \citep{Kalasetal2005}; $e_\mathrm{cav}=0.07$ for HR\,4796 \citep{Schneideretal2009,Thalmannetal2011}, and $e_\mathrm{cav}=0.18$ for HD\,202628 \citep{Kristetal2012}. In the case of Fomalhaut, recent ALMA observation at 1.3\,mm implied an eccentricity of 0.12  for the cavity and provided the first conclusive detection of apocentre glow with a apocentre-to-pericentre flux ratio of $\sim1.1$  \citep{MacGregoretal2017}.  Being bright at milimetre wavelengths \citep{Mooeretal2013} and having a confirmed $5\pm1\,M_\mathrm{Jup}$ planet \citep{Rameauetal2013}, HD\,95086 could particularly be a good target to apply our method.

\section*{Acknowledgements}

This work was supported by the the Hungarian Grant K119993 and the Momentum grant of the MTA CSFK Lend\"ulet disc Research Group. TK was supported by the Hungarian OTKA Grant No.  NK100296. AM acknowledges the support of Bolyai Research Fellowship. We gratefully acknowledge the support of NVIDIA Corporation with the donation of the Tesla 2075 and K40 GPUs. ZsR thanks to M. Wyatt and T. Pearce for helpful discussions. We also thank to the referee Alexander Mustill for thoughtful comments that helped to improve the quality of the paper.

\label{lastpage}

\end{document}